\documentclass[twocol]{ametsoc}
\journal{jas}

\usepackage{graphicx,overpic,natbib}
\usepackage{bm,url,color,colortbl,xcolor,soul}
\usepackage{url}
\usepackage{multirow}
\usepackage{booktabs} 
\usepackage[toc,page]{appendix}
\graphicspath{{./fig/}{./png/}}

\bibpunct{(}{)}{;}{a}{}{,}

\title{Effect of turbulence on collisional growth of cloud droplets}

\authors{Xiang-Yu Li\correspondingauthor{Xiang-Yu Li, Department of Meteorology and Bolin Centre for Climate Research, Stockholm University, Stockholm, Sweden}}
\affiliation{Department of Meteorology and Bolin Centre for Climate Research, Stockholm University, Stockholm, Sweden;\\
Nordita, KTH Royal Institute of Technology and Stockholm University, 10691 Stockholm, Sweden;\\
Swedish e-Science Research Centre, www.e-science.se, Stockholm, Sweden;\\
Laboratory for Atmospheric and Space Physics, University of Colorado, Boulder, CO 80303, USA;\\
JILA and Department of Astrophysical and Planetary Sciences University of Colorado, Boulder, CO 80303, USA}
\email{xiang.yu.li@su.se}
\extraauthor{Axel Brandenburg}
\extraaffil{\small
Laboratory for Atmospheric and Space Physics, University of Colorado, Boulder, CO 80303, USA;\\
Nordita, KTH Royal Institute of Technology and Stockholm University, 10691 Stockholm, Sweden;\\
JILA and Department of Astrophysical and Planetary Sciences University of Colorado, Boulder, CO 80303, USA;\\
Department of Astronomy, Stockholm University, SE-10691 Stockholm, Sweden}
\extraauthor{Gunilla Svensson}
\extraaffil{Department of Meteorology and Bolin Centre for Climate Research, Stockholm University, Stockholm, Sweden;\\
Swedish e-Science Research Centre, www.e-science.se, Stockholm, Sweden;\\
Global \& Climate Dynamics, National Center for Atmospheric Research, Boulder, CO 80305, USA}
\extraauthor{Nils E.\ L.\ Haugen}
\extraaffil{SINTEF Energy Research, 7465 Trondheim, Norway; \\
Department of Energy and Process Engineering, NTNU, 7491 Trondheim, Norway}
\extraauthor{Bernhard Mehlig}
\extraaffil{Department of Physics, Gothenburg University, 41296 Gothenburg, Sweden}
\extraauthor{Igor Rogachevskii}
\extraaffil{Department of Mechanical Engineering, Ben-Gurion Univ. of the Negev, P. O. Box 653, Beer-Sheva 84105, Israel;\\
Nordita, KTH Royal Institute of Technology and Stockholm University, 10691 Stockholm, Sweden}

\date{\today,~ $ $Revision: 1.526 $ $}

\abstract{
We investigate the effect of turbulence on the collisional growth of
$\mu$m-sized droplets through high-resolution numerical simulations with
well resolved Kolmogorov scales, assuming a collision and coalescence
efficiency of unity.
The droplet dynamics and collisions are approximated using a
superparticle approach.
In the absence of gravity,
we show that the time evolution of the shape of the droplet-size distribution
due to turbulence-induced
collisions depends strongly on the turbulent energy-dissipation rate $\bar{\epsilon}$, but
only weakly on the Reynolds number.
This can be explained through the $\bar{\epsilon}$-dependence
of the mean collision rate described by the Saffman-Turner
collision model. Consistent with the Saffman-Turner
collision model and its extensions, the collision rate increases
as $\bar{\epsilon}^{1/2}$ even when coalescence is invoked.
The size distribution exhibits power law behavior with a slope of $-3.7$
between a maximum at approximately $10\,\mu$m up to about $40\,\mu$m.
When gravity is invoked,
turbulence is found to dominate the time evolution of an initially
monodisperse droplet distribution at early times.
At later times, however, gravity takes over and dominates the collisional growth.
We find that the formation of large droplets
is very sensitive to the turbulent energy dissipation rate. This is
due to the fact that turbulence enhances the collisional growth between
similar sized droplets at the early stage of raindrop formation.
The mean collision rate grows exponentially, which is consistent
with the theoretical prediction of the continuous collisional growth even when
turbulence-generated collisions are invoked.
This consistency only reflects the mean effect of turbulence
on collisional growth.
}

\begin{document}
\maketitle

\section{Introduction}

Collisional growth of inertial particles in a turbulent environment plays
an important role in many physical processes \citep{pumir2016collisional, ohno2017condensation}.
For instance, collisional growth of droplets in atmospheric clouds
may explain the rapid warm rain formation \citep{shaw_2003, DBB12,Grabowski_2013}.
Collisions of dust grains in circumstellar disks is proposed to be a
key step towards planet formation \citep{johansen2017forming}.

The most notorious difficulty is how turbulence affects the collisional growth.
This problem has a long history and was recently reviewed by
\citet{shaw_2003}, \cite{DBB12}, \citet{Grabowski_2013} and \citet{pumir2016collisional}.
The pioneering
work by \citet{saffman_1956} proposed a theoretical model for the collision
rate (Saffman-Turner model) of cloud droplets. The key idea of the
Saffman-Turner model is that the collision rate is dominated by small
scales of turbulence since the size of cloud droplets (typical size is $10\,\mu$m in radius)
is three orders of magnitude smaller than the Kolmogorov length (i.e., the smallest scale of
turbulence, which is about 1 mm in atmospheric clouds).
The Saffman-Turner model predicts that the mean collision rate $\bar{R_c}$
is proportional to the mean energy dissipation rate $\bar{\epsilon}$
if there is no intermittency and the particle inertia is small.
Following \cite{saffman_1956}, \citet{reuter1988collection}
used a stochastic model
to show that turbulence fluctuations modeled by random perturbations enhance
the collision rate.
\citet{grover1985effect}
studied the effect of vertical turbulent fluctuations on the
collision between aerosol particles and cloud droplets using a one-dimensional model.
They inferred that three-dimensional atmospheric turbulence should cause
a substantial enhancement of the collision rate.
The stochastic model developed by \cite{pinsky2004collisions} demonstrated that
the turbulence enhancement can reach a factor of up to 5. Follow-up studies of
\cite{pinsky2007collisions, pinsky2008collisions} showed that turbulence has a significant
enhancement effect on the collision rate, especially for small and similar-sized droplets
(radius of a few $\mu$m.).
However,
\citet{koziol1996effect} found that, using a stochastic model,
turbulence only has a moderate effect on the collision rate.
This discrepancy between the two stochastic models is
either due to the simplified descriptions of the droplet motion
or an inaccurate modelling of turbulence fluctuations
\citep{wang2005theoretical, Grabowski_2013}.

Due to the rapid advances in computer technology, the collision rate
has been studied using direct numerical simulations (DNS).
Most of the DNS studies focused on the collisional growth without coalescence.
Such studies are useful in that they
facilitate our understanding of the physical mechanisms contributing to
the collision rate, such as droplet clustering and relative velocity.
\citet{sundaram97} constructed the collision rate using
the radial distribution velocity and the radial distribution function based
on the Saffman-Turner collision rate. Their subsequent work found that
turbulence has a significant effect on droplet clustering and on the
relative velocity, which demonstrated its importance for
the collision rate \citep{SRC98,CK04,CC05,SA08, ireland16_1, ireland16}.
The turbulence effect on the geometrical
collisional kernel was investigated by \citet{franklin2005collision},
\cite{ayala_2008}, \cite{rosa2013kinematic} and \citet{chen2016cloud}.
\citet{Ayala08} developed a comprehensive parameterizations
of the turbulent collision rate and concluded that turbulence increases
the geometrical collision rate by up to $47\%$ with increasing energy
dissipation rate. They also found that the dependence of the collision
rate on the Reynolds number is minor. \cite{rosa2013kinematic} and \citet{chen2016cloud}
confirmed the secondary dependency of the collision statistics on the Reynolds number.

Invoking coalescence is computationally and technically more demanding,
but more realistic.
\citet{riemer2005droplets} found that the turbulent collision rate
is several orders of magnitude larger than the gravitational
collision rate. However, \citet{wang2006comments} argued that this
result is grossly overestimated because their rms velocity was
overestimated by a factor of $\sqrt{3}$ and their treatment of
the collision efficiency only included gravity but not turbulence.
\citet{franklin2008warm} investigated collision-coalescence
processes by solving the Smoluchowski equation together with the Navier-Stokes
equation using DNS and found that the size distribution of cloud droplets
is significantly enhanced by turbulence.
Using a similar approach, \citet{xue2008growth}
concluded that even a moderate turbulence enhancement of the collision rate
can have a significant effect on the growth of similar sized droplets,
which is referred to as the auto-conversion phase of the growth.
A similar conclusion was reached by \cite{Wang_2009}, who found that
turbulence enhances the collisional growth by a factor of 2.
They also found that in their simulations the dependence on Reynolds
number is uncertain due to its small value.
\citet{2016Onishi}
updated the collision rate model of \citet{Wang_2009} and
performed DNS at higher Reynolds number. They
found that the collisional growth of cloud droplets depends on the Reynolds
number. However, they did not study the time evolution of the
size distribution nor its dependency on Reynolds number and energy
dissipation rate. \cite{chen2018turbulence} investigated the time evolution
of the size distribution and its dependence on the energy dissipation rate
and Reynolds number using a Lagrangian collision-detection method. They concluded that
turbulence has a prominent effect on the broadening of the size
distribution -- even if the turbulence intensity is small.
However, a comparison between the theoretical predictions of the
$\bar{\epsilon}$-dependence of the collision rate \citep{saffman_1956}
and the numerical simulations results was not performed
for the case when coalescence was invoked.
Also, the effect of the
initial width of the size distribution on the turbulence enhancement
was not discussed.

In fully-developed turbulence, droplet collision-coalescence process
results in a wide range of droplet sizes and thus in a wide range of droplet Stokes numbers
that evolve during the simulation.
The Stokes number is a dimensionless measure of the effect of droplet inertia, which
depends on the geometrical droplet size and the turbulence intensity.
In cloud turbulence
with $\bar{\epsilon}\approx0.04\,\rm{m}^2\rm{s}^{-3}$,
the Stokes number ${\rm St}$ varies from $10^{-3}$ (droplet radius of about $1\,\mu$m)
to $10$ (about $100\,\mu$m) and beyond.
Very small cloud droplets (for ${\rm St} \ll 1$) are advected by turbulent air flow
and the collision is caused
by local turbulent shear \citep{saffman_1956, Andersson07}.
For larger Stokes numbers, on the other hand, inertial effects become important,
that allow the droplets to detach from the flow. This may substantially increase
the collision rate \citep{Sun97,falkovich2002, Wilkinson06}.
The time-dependent collision rate due to the dynamical Stokes number cannot be captured with
a predetermined parameterization of the
collision rate.
\citet{saito2017turbulence} developed a Lagrangian
algorithm to detect collisions without using a parameterized collision kernel.
They observed that turbulence broadens the size distribution of cloud droplets.
Since their work has condensation included, it is unclear if the broadening
of the size distribution results from
the turbulence effect on the collisions or its effect on condensation.
Such a broadening of the size distribution due to condensation could result
from turbulence-facilitated supersaturation fluctuations \citep{2015_Sardina}.

To quantify the role of small-scale turbulence on the time evolution of the
size distribution and its connection to 
the Saffman-Turner collision rate \citep{saffman_1956}
when coalescence is included, 
we investigate the collisional growth of cloud droplets with or without
gravity.
We determine the droplet-size distribution directly
from numerical simulations, thus avoiding the use of a parameterized kernel.
We focus on the time evolution of the size distribution due to
collision-coalescence processes and
how changing the Reynolds number and the energy dissipation
rate affect the size distribution.
We perform high resolution DNS
of turbulence with a well resolved Kolmogorov viscous scale
(our maximum Taylor-microscale Reynolds number
is $158$).
Droplet and collision dynamics are solved together using a superparticle approach
assuming unit collision and coalescence efficiency. Unit coalescence efficiency means
that droplets coalesce upon collision.
Compared with the direct Lagrangian collision-detection method, the superparticle
approach is computationally less demanding.
This can be deduced from a cross comparison with the Eulerian approach.
First, \citet{li17} found that the superparticle is about 10 times faster than
an Eulerian approach where one solves
the Smoluchowski and momentum equations for logarithmically spaced mass bins.
Second, the direct Lagrangian collision-detection method is more costly than
the Eulerian approach \citep{chen2018turbulence}.
Therefore, the superparticle approach is at least 10 times faster than the
direct Lagrangian collision-detection method.
More importantly, the superparticle approach can easily be extended to large
eddy simulations with an appropriate sub-grid scale model \citep{grabowski2017broadening}.
A detailed comparison of the present simulation results and the theoretical prediction
of the collision rate \citep{saffman_1956} is conducted.
In addition, we explore how the width of the initial size distribution
alters the turbulence effect on collisional growth of cloud droplets.
In the meteorology community, the process
of collision and coalescence is referred to as collection, while in the astrophysical
community, this process is referred to as coagulation. Since we assume unit coalescence
efficiency in the present study, we use the terminology collision.
To address the turbulence-facilitated collision
for more general applications (such as interstellar dust),
we will first focus on DNS without gravity. We will then
turn to DNS with both gravity and turbulence, which is
important for the cloud droplet formation.

\section{Numerical setup}
\label{NumericalSetup}

\begin{table*}[t!]
\caption{Summary of the simulations.}
\centering
\setlength{\tabcolsep}{3pt}
\begin{tabular}{lccccccccr}
Run & $N_p/10^6$ & $N_{\rm grid}$ &  $f_0$ & L (\rm{m}) & $u_{\rm rms}$ ($\rm{m} \, \rm{s}^{-1}$) & $\mbox{\rm Re}_{\lambda}$ & $\bar{\epsilon}$ ($\rm{m}^2\rm{s}^{-3}$)& $\eta\cdot 10^{-4}$ ($\rm{m}$) & $\tau_{\eta}$ ($\rm{s}$)\\
\hline
A & $8.4$ & $256^3$ & $0.02$    & 0.125 &0.17 &57  & 0.039 & 4  & 0.016 \\ 
B & $67$  & $512^3$ & $0.02$    & 0.25  &0.21 &94  & 0.04  & 4  & 0.016 \\ 
C & $67$  & $512^3$ & $0.02$    & 0.50  &0.27 &158 & 0.036 & 4  & 0.017 \\ 
D & $67$  & $512^3$ & $0.0072$ & 0.44  &0.13 &98  & 0.005 & 7  & 0.044 \\ 
E & $67$  & $512^3$ & $0.01$    & 0.37  &0.15 &97  & 0.01  & 6  & 0.032 \\ 
F & $67$  & $512^3$ & $0.014$  & 0.30  &0.18 &94  & 0.02  & 5  & 0.022 \\ 
\hline
\multicolumn{10}{p{0.65\textwidth}}{
Here, $f_0$ is the amplitude of the random forcing (see text) and $L$ is the domain size.
}
\end{tabular}
\label{Swarm_Rey}
\end{table*}

Our simulations are conducted using the {\sc Pencil Code}.
The DNS of the turbulent flow are performed
for a weakly compressible gas, and we adopt a superparticle algorithm to approximate
the droplet dynamics \citep{Dullemond_2008,Shima09,Johansen_2012}.

{\em DNS of the turbulent air flow}.
The velocity $\bm{u}$ of the turbulent air flow is determined by
the Navier-Stokes equation:
\begin{equation}
	{\partial{\bm u}\over\partial t}+\bm{u}\cdot{\bm{\nabla}}\bm{u}={\bm f}
-\rho^{-1}{\bm{\nabla}} p
	+\rho^{-1} {\bm \nabla} \cdot (2 \nu \rho \bm{{\sf S}}) ,
\label{turb}
\end{equation}
where ${\bm f}$ is a monochromatic random forcing function \citep{Brandenburg01},
$\nu$ is the kinematic viscosity of the air flow,
$\bm{{\sf S}}_{ij}={\textstyle\frac{1}{2}}(\partial_j u_i+\partial_i u_j)
-{\textstyle{1\over3}}\delta_{ij}{\bm{\nabla}}\cdot\bm{u}$ is
the traceless rate-of-strain tensor,
$p$ is the gas pressure, and $\rho$ is the gas density,
which in turn obeys the continuity equation,
\begin{equation}
	{\partial\rho\over\partial t}+{\bm{\nabla}}\cdot(\rho\bm{u})=0 .
\end{equation}
We assume that the gas is isothermal with constant sound speed $c_{\rm s}$,
so that $c_{\rm s}^2=\gamma p/\rho$, where $\gamma=c_{\rm P}/c_{\rm V}=7/5$
is the ratio between specific heats, $c_{\rm P}$ and $c_{\rm V}$, at constant
pressure and constant volume, respectively.
To avoid global transpose operations associated with calculating
Fourier transforms for solving the nonlocal equation for the pressure in strictly
incompressible calculations, we solve here instead the compressible Navier-Stokes
equations using high-order finite differences.
To simulate the nearly incompressible atmospheric air flow,
we set the sound speed to $5\,\rm{m} \, \rm{s}^{-1}$,
resulting in a Mach number
of $0.06$ when the $u_{\rm rms}=0.27\,\rm{m} \, \rm{s}^{-1}$.
Such a configuration with so small Mach number is equivalent to an incompressible flow.
Indeed, we quantify the weak compressibility in our DNS by
calculating the dimensionless number
$\wp=\textstyle{\left<\left|\bm{\nabla}\cdot\bm{u}
\right|^2\right>/\left<\left|\bm{\nabla}\times\bm{u}\right|^2\right>
=2\times10^{-4}}$.
Following \citet{GM16},
the parameter $\wp=2\times10^{-4}$ corresponds
to a Stokes number ${\rm St}=0.018$.
The smallest Stokes number in our DNS is ${\rm St}=0.05$.
This implies that the effect of fluid compressibility is
much less than the compressibility of the particle velocity field
caused by droplet inertia.
Therefore, the effect of fluid compressibility on the droplets is also negligible.

To characterize the intensity of turbulence, we use the Taylor microscale Reynolds number
$\mbox{\rm Re}_{\lambda} \equiv u_{\rm rms}^2 \sqrt{5/(3\nu\bar{\epsilon})}$,
where $u_{\rm rms}$ is the rms turbulent velocity,
and
$\bar{\epsilon}=2\nu\, \textstyle{\overline{{\rm Tr\,} {\sf S_{ij}} {\sf S_{ji}}}}$
is the mean energy-dissipation rate per unit mass and $\rm{Tr}$ denotes the trace.
The parameters of all simulations are listed in Table~\ref{Swarm_Rey}.
Here $\tau_{\eta}=\left(\nu/\bar{\epsilon}\right)^{1/2}$ is the Kolmogorov time and
$\eta=\left(\nu^3/\bar{\epsilon}\right)^{1/4}$ is the Kolmogorov length.

{\em Superparticle algorithm}.
The equations governing the dynamics
and collision of droplets in a turbulent flow are solved
simultaneously with the Navier-Stokes equations.
We approximate the droplet dynamics and collisions using a stochastic Monte Carlo algorithm
\citep{bird1978monte,bird1981monte,jorgensen1983comparison}
that represents a number of spherical droplets by a superparticle
\citep{Dullemond_2008, Shima09, Johansen_2012, li17}.
All droplets in superparticle $i$ are assumed to have the same
material density $\rho_d$, radius $r_i$, and velocity $v_i$.
Further, each superparticle is assigned a volume of the grid cell and thus
a droplet number density, $n_i$.
The position
$\bm{x}_i$ of superparticle $i$ is determined by
\begin{equation}
	\frac{d\bm{x}_i}{dt}=\bm{V}_i
\label{dxidt}
\end{equation}
and
\begin{equation}
	\frac{d\bm{V_i}}{dt}=\frac{1}{\tau_i}(\bm{u}
	-\bm{V}_i) + \bm{g}\,.
\label{dVidt}
\end{equation}
Here,
\begin{equation}
\label{response_time}
\tau_i=2\rho_{\rm d} r_i^2/[9\rho\nu \, D({\rm Re}_i)]
\end{equation}
is the particle response time attributed to superparticle $i$.
The correction factor \citep{Schiller33, Marchioli08}
\begin{equation}
D(Re_i)=1+0.15\,{\rm Re}_i^{2/3}
\label{eq:correction}
\end{equation}
models the effect of non-zero particle Reynolds number
${\rm Re}_i=2r_i|\bm{u}-\bm{V}_i|/\nu$.
This is a widely used approximation, although it does not
correctly reproduce the small-$\rm{Re}_i$ correction to Stokes formula \citep{John07}.
The dimensionless particle-response time is given by the Stokes number ${\rm St}=\tau_i/\tau_{\eta}$.
Droplets are randomly distributed in the simulation domain with zero velocity initially.
The term $\bm{g}$ in equation~\eqref{dVidt} is included only
when collisions are also driven by gravity, in addition to turbulence.

Droplet collisions are represented by collisions of superparticles
\citep{Shima09,Johansen_2012,li17}.
When two superparticles collide, two droplets in either of
the superparticles can collide with probability
$p_{c}=\tau_{c}^{-1}\Delta t$,
where $\Delta t$ is the integration time step.
A collision event occurs when $p_{c}>\eta_c$, where $\eta_c$ is a random number.
If a collision
event happens, $\eta_c$ must lie between zero and one; see appendix~\ref{collision}
for details on the collision scheme.
A mean-field model is adopted for the collision time $\tau_{c}$:
\begin{equation}
	\tau_{c}^{-1}=\sigma_{c} n_j\left|\bm{V}_{i}
	-\bm{V}_{j}\right| E_{c}.
\label{tauij1}
\end{equation}
Here $\sigma_{c}=\pi(r_i+r_j)^2$
is the geometric collision cross section between two droplets with radii $r_i$  and $r_j$.
The parameter $E_{c}$ is the collision efficiency \citep{DBB12}.
It is set to unity in our simulations, and we assume that droplets coalesce upon collision.
The use of a unit collision efficiency overestimates the collision rate.
This is done to reduce the complexity of the simulation
and enables us to focus only on the collision dynamics. Adopting unit coalescence
efficiency is justified by the fact that the Weber number is only about $10^{-2}$
\citep{perrin2015lagrangian}.
The Weber number is defined as the ratio of droplet inertia and
its surface tension \citep{perrin2015lagrangian}.
A low Weber number means that the colliding droplets always coalescence.
It is worth noting that a cylindrical kernel is used in the present superparticle scheme
as described by equation~\eqref{tauij1}.
This is in contrast to what is done for Lagrangian point particle simulations \citep{wang1998statistical}.
In those simulations, a spherical kernel was used, where collisions were not enabled.
The use of a cylindrical kernel in the superparticle approach is justified because
the superparticle approach treats collisions in a statistical fashion,
where the interacting superparticles are considered to fill the entire grid cell.

Collisions are enabled at the same time when the simulation starts with $\bm{u}=\bm{0}$.
This yields virtually the same result compared to the case when
turbulence is already fully developed and droplets
are mixed; see appendix~\ref{app:initialCondition}.
Since collisions can only happen when a pair of superparticles resides
in the same grid cell, it is important to have sufficient statistics of initial $N_{\rm p}/N_{\rm grid}$
($N_{\rm p}$ is the number of superparticles and $N_{\rm grid}$ is the number of grid cells).
Furthermore, to obtain
fully developed turbulence, a large number of mesh points $N_{\rm grid}$ ($512^3$ in the present study)
is essential. This requires a large number of superparticles,
which is computationally expensive even on the modern supercomputers.
We investigate the statistical convergence with respect to the initial value of $N_{\rm p}/N_{\rm grid}$
and find that it is converged at $0.05$ (see appendix~\ref{app:Np/Ngrid}).
Nevertheless, to have sufficient droplet statistics, we adopt $N_{\rm p}/N_{\rm grid}=0.5$,
so we have on average one superparticle for every two grid cells.
This makes the computation affordable since the computational cost
scales as $N_{\rm p}^2$.
Droplet growth by condensation is not incorporated in our model.
We refer to \citet{li17} for a detailed description of our numerical
setup and of the algorithm used to model collision.

The superparticle approach is computationally efficient \citep{li17,Shima09, Johansen_2012},
but it is an approximation.
How accurately it describes the actual microscopic
collision dynamics depends on several factors.
In the limit where the number of droplets
per superparticle tends to infinity, the algorithm reduces to a full mean-field description
\citep{Dullemond_2008, 1997_pruppacher}.
In the opposite limit, when the number of droplets per superparticle is small,
the algorithm incorporates fluctuations in the collision processes that may be important
in the dilute system that we consider here \citep{Kos05,wilkinson16}.
\citet{Dziekan17} compared the superparticle approach with the direct detection
of collisions of point particles and concluded that the superparticle approach can
accurately describe such fluctuations as long as
the number of droplets is below 10 per superparticle. In our simulations,
we assign two droplets per superparticle to ensure that the algorithm is sufficiently accurate.

In our simulations, we check for collisions at each time step, which enables us to get the
size distribution $f(r,t)$ at time $t$ and droplet radius $r$.
This distribution not only determines
rain formation in clouds, but also the optical depth of the cloud \citep{Beals15}.

{\em Initial conditions}.
As initial condition, we adopt a log-normal droplet-size distribution
\citep{nenes2003parameterization, Sei06}
is widely used in climate models and
is supported by the in situ atmospheric measurements \citep{miles2000cloud},
\begin{equation}
\label{init_dist}
f(r,0)={n_0 \over \sqrt{2\pi}\sigma_{\rm ini} \, r} \, \exp\Big[
-{\ln^2 (r/r_{\rm ini}) \over 2\sigma_{\rm ini}^2} \Big]\,.
\end{equation}
Here $r_{\rm ini}=10\,\mu$m
and $n_0=n(t=0)$ is the initial
number density of droplets.

To speed up the computation by a factor of a hundred,
we adopt $n_0=10^{10} \, \rm{m}^{-3}$ instead of the typical
value in the atmospheric clouds, $n_{\rm ref}\equiv10^8 \, \rm{m}^{-3}$; cf.\ \citet{li17}.
We explore the convergence of $\sigma_{\rm ini}$ for collision driven by
combined turbulence and gravity. It is found that $\sigma_{\rm ini}$ converges
at $0.02$ (see appendix~\ref{app:width}). However, since gravity-generated
collision is very sensitive to the
initial size difference, we employ monodisperse initial distribution ($\sigma_{\rm ini}=0$) for
the case of combined turbulence and gravity.
For turbulence-generated collision without gravity, we employ $\sigma_{\rm ini}=0.2$.

\section{Results and discussion}

\subsection{Collisions driven by turbulence}

\begin{figure*}[t!]
\begin{center}
\includegraphics[width=\textwidth]{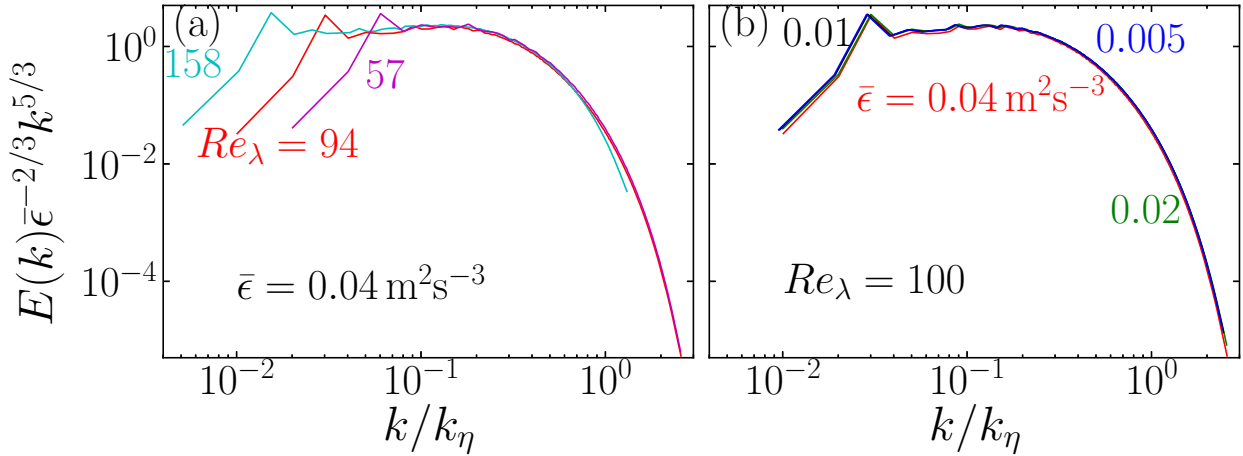}
\end{center}
\caption{Time-averaged kinetic energy spectra of the turbulence gas flow
for (a) different $\mbox{\rm Re}_{\lambda}$ = 57 (magenta dashed line),
94 (red solid line), and 158 (cyan dotted line) at fixed $\bar{\epsilon}$ = $0.04\rm{m}^2\rm{s}^{-3}$
(see Runs A, B, and C in Table~\ref{Swarm_Rey} for details)
and for (b) different $\bar{\epsilon}$ = $0.005\,\rm{m}^2\rm{s}^{-3}$ (blue dotted line),
0.01 (black dashed line), 0.02 (green dash-dotted line) and 0.04
(red solid line) at fixed $\mbox{\rm Re}_{\lambda}=100$
(see Runs B, D, E, and F in Table~\ref{Swarm_Rey} for details).
}
\label{ppower_FixE0p03}
\end{figure*}

Figure~\ref{ppower_FixE0p03}(a) shows
the time-averaged turbulent kinetic-energy spectra
for different values of $\mbox{\rm Re}_{\lambda}$ at fixed
$\bar{\epsilon}\approx0.04\,\rm{m}^2\rm{s}^{-3}$.
Here, $\mbox{\rm Re}_{\lambda}$ is varied by changing
the domain size $L$, which in turn changes $u_{\rm rms}$.
For
larger Reynolds numbers the spectra extend to smaller wavenumbers.
Since the energy spectrum is compensated by
$\bar{\epsilon}^{-2/3}k^{5/3}$,
a flat profile corresponds to
Kolmogorov scaling \citep{Pope00}.
For the largest $\mbox{\rm Re}_{\lambda}$ in our simulations
($\mbox{\rm Re}_\lambda=158$),
the inertial range extends for about a decade
in $k$-space.
Figure~\ref{ppower_FixE0p03}(b) shows how the energy spectra depend on $\bar{\epsilon}$.
Here we keep the values of $\mbox{\rm Re}_{\lambda}$ and
$\nu$ fixed, but vary $u_{\rm rms}$
by changing both $L$ and the amplitude of the forcing; see Table~\ref{Swarm_Rey} for details.
Since the abscissa in the figures is normalized by $k_\eta=2 \pi /\eta$,
the different spectra shown in Figure~\ref{ppower_FixE0p03}(b) collapse onto a single curve.

\begin{figure*}[t!]
\begin{center}
\includegraphics[width=\textwidth]{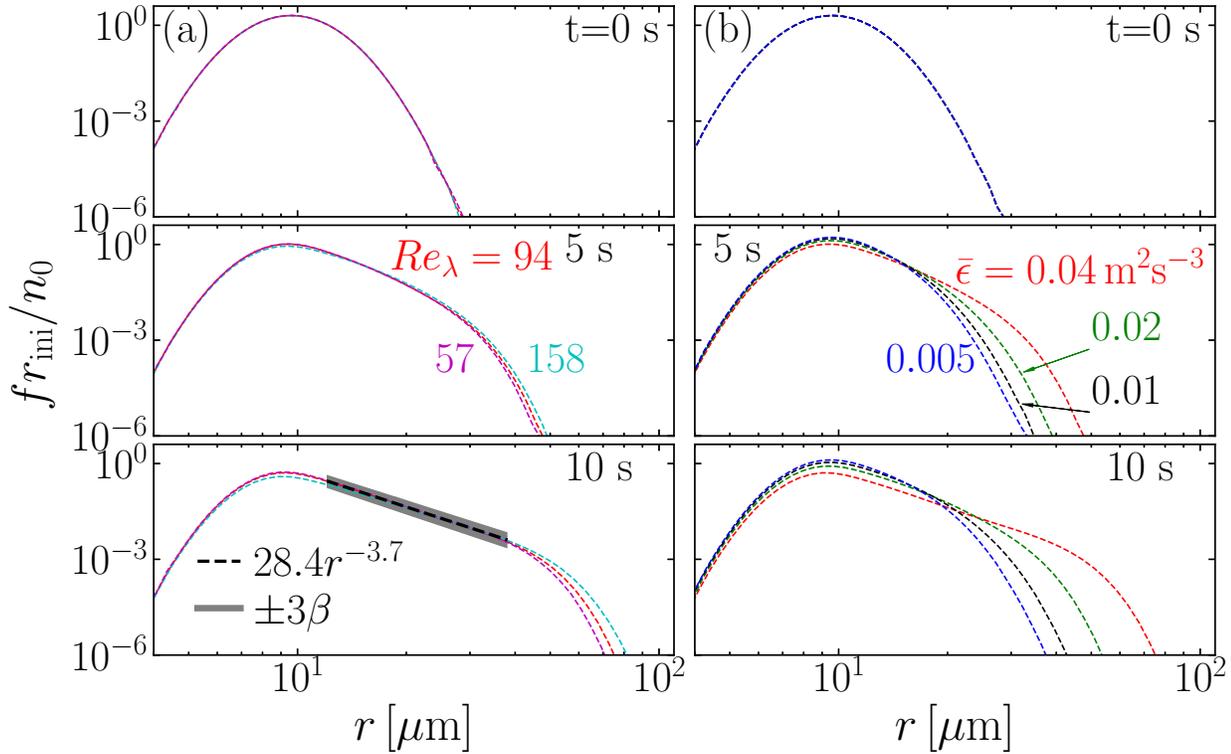}
\end{center}
\caption{Droplet-size distribution for the same simulations as in Figure~\ref{ppower_FixE0p03}.
(a) Different ${\rm Re}_\lambda$ at fixed $\bar{\epsilon}$.
(b) Different $\bar{\epsilon}$ at fixed ${\rm Re}_\lambda$.
Here $\beta$ is the standard deviation and $3\beta$ is the significance level.
}
\label{f_shima_grav0Re100_cond0_coa_comp}
\end{figure*}

Figure~\ref{f_shima_grav0Re100_cond0_coa_comp}(a) shows the droplet-size distributions
obtained in our simulations for different values of $\mbox{\rm Re}_{\lambda}$,
but for the same $\bar{\epsilon}$.
This figure demonstrates that the time evolution of the size
distribution depends only weakly on $\mbox{\rm Re}_{\lambda}$ when $\bar{\epsilon}$ is kept constant.
This is consistent with the notion
that the collisional growth is mainly dominated by the Kolmogorov scales  \citep{saffman_1956,DBB12}.
The maximum Reynolds number in our DNS is
$\mbox{\rm Re}_{\lambda}\approx 158$.
This value is still two orders of magnitude smaller than the
typical value in atmospheric clouds \citep{Grabowski_2013}.
It cannot be ruled out that there may
be a stronger Reynolds-number effect on the collisional growth
at higher Reynolds numbers \citep{shaw_2003,ireland16_1,2016Onishi}.
In the simulations of \citet{2016Onishi}, where collisions are detected directly,
the largest value of ${\rm Re}_{\lambda}$ was $333$, which is twice
as large as our largest value. They showed that the turbulence enhancement factor
weakly depends on ${\rm Re}_{\lambda}$ when
the mean radius of the initial distribution is
$10\,\mu$m.
This is consistent with our results.

Figure~\ref{f_shima_grav0Re100_cond0_coa_comp}(b) shows how
the evolution of the droplet-size
distribution depends on $\bar{\epsilon}$, for a fixed $\mbox{\rm Re}_{\lambda}$.
We see that especially
the tails of the size distributions depend strongly on $\bar{\epsilon}$:
the larger $\bar{\epsilon}$, the wider are the tails.
The tails in the droplet-size distribution lead
to a broad distribution of Stokes numbers.
Also, since ${\rm St}=\tau_i/\tau_{\eta}\propto \bar{\epsilon}^{1/2}$,
the ${\rm St}$-distribution shifts to large Stokes numbers as $\bar{\epsilon}$
increases (see appendix~\ref{app:StokesNumner}).

\begin{figure*}[t!]
\begin{center}
\includegraphics[width=\textwidth]{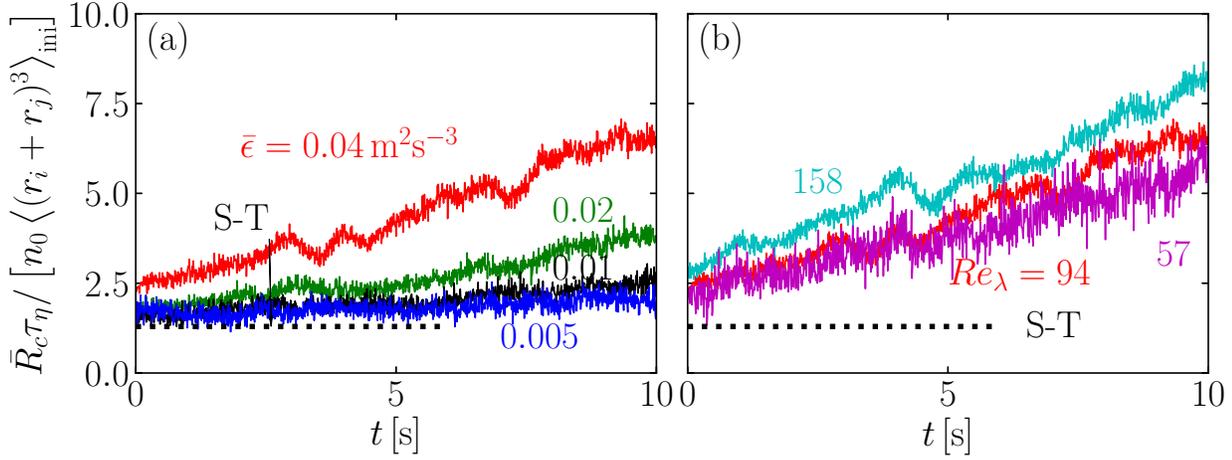}
\end{center}
\caption{Mean collision rate for same simulations as in Figure~\ref{ppower_FixE0p03}.
Panel (a): different $\bar{\epsilon}$ at fixed ${\rm Re}_\lambda$.
Panel (b): different ${\rm Re}_\lambda$ at fixed $\bar{\epsilon}$.
In both panels the data are normalized by dividing by $n_0\left\langle(r_i+r_j)^3\right\rangle_{\rm ini}/\tau_{\eta}$.
}
\label{NCOAGPM_grav0Re100_cond0_coa_comp}
\end{figure*}

We now show that the $\bar{\epsilon}$-dependence of the size distribution
is due to the sensitive dependence of the collision rate upon this parameter.
Figure~\ref{NCOAGPM_grav0Re100_cond0_coa_comp} shows
how the mean collision rate $\overline{R}_{\rm c}$ changes as a function of time.
This rate, which depends implicitly on $\bar{\epsilon}$, is defined as
\begin{equation}
\overline{R}_{\rm c} = \pi n_0 (2r)^2 \overline{|\bm{v}|},
\end{equation}
where $\bm{v}$ is the relative velocity between two approaching droplets.
This expression is written for identical droplets with radius $r$.
In bidisperse suspensions with droplets of two different radii $r_i$ and $r_j$, $2r$ is replaced by $r_i+r_j$.
Collisions of small droplets advected by turbulence are due to local turbulent
shear, provided that droplet inertia is negligible.
\citet{saffman_1956} proposed an expression for the resulting collision rate:
\begin{equation}
\label{eq:ST}
R_{\rm c}^{\rm S. T.} =  \frac{C \,n_0\, (2r)^3}{\tau_{\eta}}\,.
\end{equation}
\citet{saffman_1956} quote the value $C=\sqrt{8\pi/15}\approx1.29$ for the prefactor,
but this is just an approximation, even at
${\rm St}=0$~\citep{Wilkinson14}. It turns out that the Saffman-Turner estimate is an upper bound
\citep{gustavsson2016statistical},
because it counts recollisions that must not be counted when the droplets coalesce upon collision,
as in our simulations.
Here recollision means that one droplet can experience several collision
since there is no coalescence.
DNS of small droplets in turbulence
also count recollisions (no coalescence) and yield a value  of $C$ in good agreement
with the Saffman-Turner estimate \citep{Wilkinson14}, in the limit of ${\rm St}\to 0$.

In Figure~\ref{NCOAGPM_grav0Re100_cond0_coa_comp}(a) we normalized
the mean collision rate by dividing with the Saff\-man-\-Tur\-ner expression (\ref{eq:ST})
for the collision rate,  averaging $(2r)^3=(r_i+r_j)^3$ over the initial size distribution.
In this way, we obtain the coefficient $C$ from the output of the mean collision
rate $\bar{R}_c$.
Initially the collision rate is of the same order as predicted by Eq.~(\ref{eq:ST}), but
in our simulations the coefficient $C$ depends on $\epsilon$.
It ranges from $C\approx 1.57$ at $\bar{\epsilon}=0.005\,\rm{m}^2\rm{s}^{-3}$
to $C\approx 2.26$ at $\bar{\epsilon}=0.04\,\rm{m}^2\rm{s}^{-3}$.
All values are larger than the Saffman-Turner prediction
in spite that the Saffman-Turner collision rate is argued to be an
upper bound for advected droplets \citep{Gustavsson08}.
However, in our simulations the mean Stokes number ranges from ${\rm St}=0.05$
for $\bar{\epsilon}=0.005\,\rm{m}^2\rm{s}^{-3}$ to ${\rm St}={0.14}$ for
$\bar{\epsilon}=0.04\,\rm{m}^2\rm{s}^{-3}$.
From Figure~1 of \citet{Wilkinson14} we infer that
$C=1.9$ for ${\rm St}=0.05$, in reasonable agreement with our simulation results.
However, their $C=5$ for ${\rm St}=0.14$, which is about twice as large
as our value ($C\approx 2.26$). This overestimation of $C$ at ${\rm St}=0.14$
could be due to their recollisions. We conclude that the collision rate
scales initially as predicted by the Saffman-Turner theory,
$R_c \sim \sqrt{\bar{\epsilon}}$, with small corrections due to particle inertia.
At later times these corrections become larger.
In recent years, several works have indicated that the Saffman-Turner
model underestimates the collision rate at larger Stokes numbers
when the effect of droplet inertia becomes important, so that the
droplets can detach from the flow. Model calculations show that
this can substantially enhance the collision rate. Two mechanisms have been proposed.

First, droplet inertia causes identical droplets to cluster spatially
\citep{M87,Elperin96,Elperin02,2000_Collins,2001_Shaw,Bec03,Dun05,EKLR13,GM16}.
At small spatial scales the clustering of identical droplets is fractal. This enhances  the collision rate
of small droplets
\citep{Gustavsson08}:
$
R_{\rm c}=  {C\,n_0\, (2r)^3}\tau_{\eta}^{-1}\,g(2r)
$.
Here $g(2r)$ is the pair correlation function measuring the degree of fractal clustering of identical droplets: $g(2r)$ diverges
$\sim r^{-\xi}$ as $r\to 0$ with $\xi>0$. The exponent $\xi$ has been computed
in DNS and model calculations \citep{GM16}. It has a weak dependence
on $\bar{\epsilon}$.
However, $g(2r)$ is calculated based on the particle field with a single Stokes number,
which makes it impossible to attempt a quantitative comparison between this theory and our simulation data.
More importantly, collision leads to a distribution of droplet sizes.
Droplets of different sizes cluster onto different fractal attractors.
This may reduce the effect of spatial clustering on the collision rate \citep{Chu05,Bec05,M17}.

Second, singularities in the droplet dynamics (caustics) give rise to multi-valued droplet velocities,
resulting in  large velocity differences between nearby droplets
\citep{Sun97,falkovich2002,Wilkinson06,FP07,Gustavsson13,Wilkinson14}.
Most model calculations were performed for identical droplets. They indicate that
the enhancement of the collision rate due to multi-valued droplet velocities dominates
for Stokes numbers larger than unity \citep{Wilkinson14}. In this case,
a Kolmogorov-scaling argument suggests \citep{Meh07,Gus08b} that
$R_{\rm c} \sim n_0 r^2 u_{\rm K} \sqrt{{\rm St}}\propto \bar{\epsilon}^{1/2}$,
where $u_{\rm K}$
is the turbulent velocity at the Kolmogorov scale;
this $\sqrt{\rm St}$-dependence was first suggested by \citet{Vol80} using a different argument.
This expression has the same $\bar{\epsilon}$-dependence as Eq.~(\ref{eq:ST}).
We note, however, that the Kolmogorov-scaling argument leading to
this $\sqrt{\rm St}$-dependence rests on the assumption that there
is a well developed inertial range (${\rm Re}_\lambda\to\infty$). This assumption is not fulfilled in our simulations.
Moreover, at later times we expect that collisions between droplets of different sizes make an
important contribution \citep{M17}. Scaling theory \citep{Meh07} suggests that the $\bar{\epsilon}$-scaling remains the same in the
limit of ${\rm Re}_\lambda\to\infty$. But, again, this limit is not realized in our simulations.
Also, any theory for the collision rate in bidisperse suspensions must be averaged over
the distribution of particle sizes and their velocities to allow comparison
with Figure~\ref{NCOAGPM_grav0Re100_cond0_coa_comp}(a). This may introduce additional
$\bar{\epsilon}$-dependencies.
It is therefore plausible that the small-${\rm St}$ scaling,
$R_c \sim \sqrt{\bar{\epsilon}}$, breaks down in our simulations at larger Stokes numbers,
indicating that the increase in the mean collision rate could be an inertial effect.
Moreover, since the Stokes numbers are larger for larger values of
$\bar{\epsilon}$, we expect the inertial additive corrections to the collision
rate (due to clustering and increased relative particle velocities) to be larger at larger $\bar{\epsilon}$. This is consistent
with Figure~\ref{NCOAGPM_grav0Re100_cond0_coa_comp}(a).
In conclusion, the mean collision rate depends strongly on $\bar{\epsilon}$ (Figure~\ref{NCOAGPM_grav0Re100_cond0_coa_comp}(a)),
as do the size distributions shown in
Figure~\ref{f_shima_grav0Re100_cond0_coa_comp}(b).

Figure~\ref{NCOAGPM_grav0Re100_cond0_coa_comp}(b) shows that the mean collision
rate depends only weakly on the Reynolds number. It demonstrates that the collision rate
is somewhat larger for larger Reynolds numbers. This is consistent with the notion that particle pairs
exploring the inertial range collide at larger relative velocities when the inertial range is larger \citep{Gus08b}. But, as pointed out above,
the inertial range in our simulations is too small for this mechanism to have a substantial effect.

It is interesting to note that the size distribution
exhibits power law behavior
in the range of $10\sim40\,\mu$m, as shown in the third panel of
Figure~\ref{f_shima_grav0Re100_cond0_coa_comp}(a).
A slope of $-3.7$ is observed.
Remarkably,
similar power laws have been observed in several other circumstances,
where the collisional growth is not subjected to gravity.
First, the observed size distribution of interstellar dust grains
shows a power law with a slope of $-3.3...-3.6$ \citep{mathis1977size}.
The collisional growth of such dust grains
in a turbulent environment is one of the main mechanisms for planet
formation \citep{johansen2017forming}.
Another example concerns the
size distribution of particles in Saturn's rings
\citep{brilliantov2015size}, where a slope of $-3$ is observed.
This power law size distribution may be universal for turbulence-generated collisional
growth.
Therefore, turbulence-generated collisional growth
without or with weak gravity
is relevant to other applications.
Next, to understand the warm rain formation, we compare with the case where gravity is included.

\subsection{Collisions driven by combined turbulence and gravity}
\label{gt}

\begin{figure}[t!]\begin{center}
\includegraphics[width=0.5\textwidth]{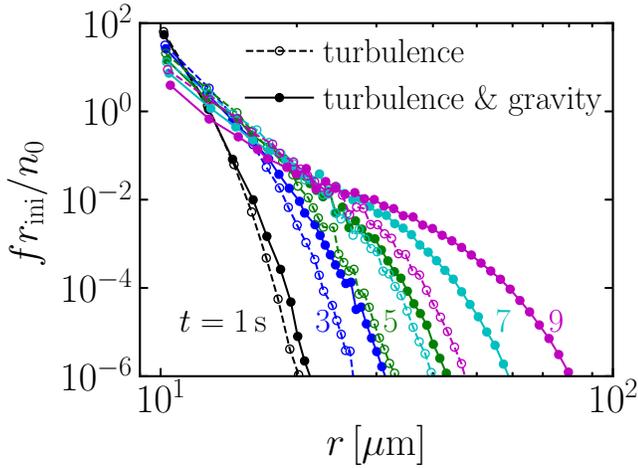}
\end{center}
\caption{Time evolution of the droplet-size distribution.
Comparing pure turbulence case (open symbols) with turbulence \& gravity
case (filled symbols). The time interval is $2\,$s, plotted from $1\,$s to $9\,$s (from left to right).
The mean energy dissipation rate is $\bar{\epsilon}=0.04\,\rm{m}^2\rm{s}^{-3}$ and
$\rm{Re}_{\lambda}=100$. The droplets are all of size $10\,\mu$m initially;
see Run D in Table~\ref{Swarm_Rey} for details of the simulation.}
\label{f_lucky_comp}
\end{figure}

For cloud-droplet growth, gravitational settling is significant
\citep{Woittiez09,Grabowski_2013}.
Collision driven by gravity is very sensitive to the initial
size difference (see appendix~\ref{app:width}). To avoid any bias
from the initial size difference, we adopt a monodisperse initial
distribution, i.e., $\sigma_{\rm ini}=0$.
In Figure~\ref{f_lucky_comp}, we compare the evolution of the
droplet-size distribution for the turbulent case and
the combined case with turbulence and gravity. At $t=1\,\rm{s}$, both
cases have almost the same droplet-size distribution,
demonstrating that turbulence dominates the collisional growth.
When $t\ge1\,\rm{s}$, gravity dominates
the time evolution of the droplet-size distribution.
The tail of the droplet-size distribution reaches
$80\,\mu$m (drizzle-size) for the combined
turbulence and gravity case at $t=9\,\rm{s}$.
For the turbulence case, the tail reaches $48\,\mu$m
after the same time, which is roughly half the radius obtained for
the combined turbulence and gravity case.
Since our initial number density of cloud droplets is a hundred times larger than the
typical value in atmospheric clouds, we can scale our simulation time by a factor of a hundred.
Thus, a scaled time of $9\,$s, for example, corresponds to $900\,$s in atmospheric clouds.
This rescaling is validated
in appendix~\ref{rescaling}, where the tail of the size distribution is found to differ by only $5\mu$m
in radius for $n_0=10^{10}\,\rm{m}^{-3}$ and $n_0=10^8\,\rm{m}^{-3}$.
We find that collisional growth of cloud droplets can reach drizzle-sized
droplets in about $900\,$s$=15\,$min.
This is comparable to the time scale for rapid warm rain formation.
This time scale, however, is expected to change if a
turbulence-induced collision efficiency were to be taken into account.
Besides, the cloud system
is about 100 times more dilute in particle number density, which may
also change the time scale.

\begin{figure}[t!]\begin{center}
\includegraphics[width=0.5\textwidth]{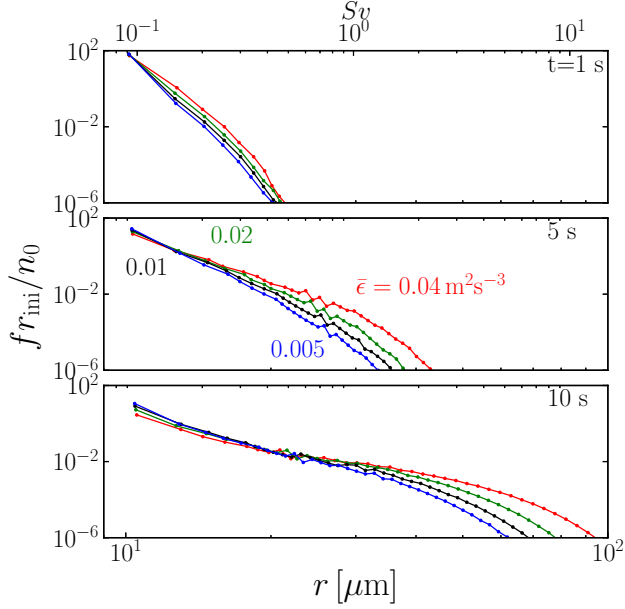}
\end{center}
\caption{Time evolution of the droplet-size distribution for different
$\bar{\epsilon}$ in combined turbulence \& gravity environment.
$Re_{\lambda}=100$. Droplets are all with size $10\,\mu$m initially.
See Runs A, B, C, and D in Table~\ref{Swarm_Rey} for details of the simulations.}
\label{f_lucky_gravRe100_cond0_coa}
\end{figure}

\begin{figure}[t!]
\begin{center}
\includegraphics[width=0.5\textwidth]{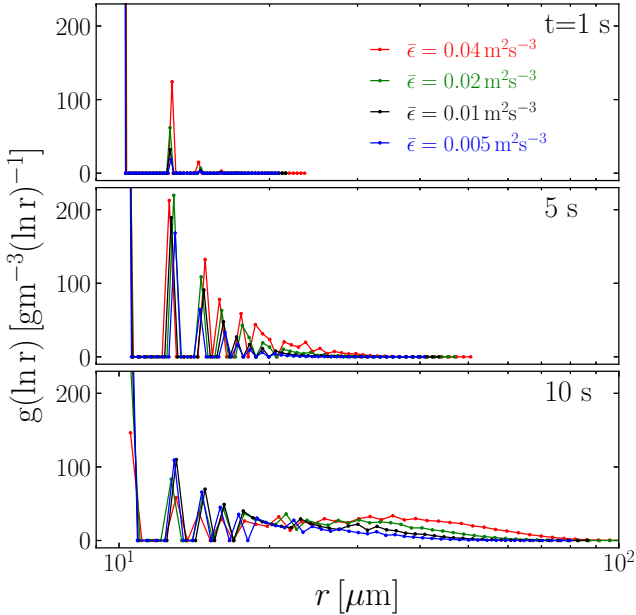}
\end{center}
\caption{Time evolution of the mass distribution function ${\rm g}(\ln{r},t)$.
Same simulations as in Figure~\ref{f_lucky_gravRe100_cond0_coa}. The values of the peaks
at $t=1$ s are $18.39$ ($\bar{\epsilon=0.005\,\rm{m}^2\rm{s}^{-3}}$),
$31.92$ ($\bar{\epsilon=0.01\,\rm{m}^2\rm{s}^{-3}}$),
$61.83$ ($\bar{\epsilon=0.005\,\rm{m}^2\rm{s}^{-3}}$),
$124.26$ ($\bar{\epsilon=0.005\,\rm{m}^2\rm{s}^{-3}}$), respectively}
\label{m_lucky_gravRe100_cond0_coa_less}
\end{figure}

Next, we check the $\bar{\epsilon}$ dependency
for the
combined turbulence \& gravity case.
As shown in Figure~\ref{f_lucky_gravRe100_cond0_coa},
the tail of the size distribution broadens with increasing
$\bar{\epsilon}$. When $fr_{\rm ini}/n_0=10^{-6}$ at $t=10\,$s, the radius
resulting from $\bar{\epsilon}=0.04\,\rm{m}^2\rm{s}^{-3}$ is about $60\,\mu$m,
while the one resulting from $\bar{\epsilon}=0.005\,\rm{m}^2\rm{s}^{-3}$
is about $90\,\mu$m. The $30\,\mu$m difference means turbulence efficiently
enhance the collisional growth when $\sigma_{\rm ini}=0$.
To quantify the role of turbulence at different phases during the collisional growth,
we inspect the mass distribution function \citep{berry_1974}.
We use the same nomenclature of the mass distribution as \citet{berry_1974},
$g(\ln{r},t)$. The mean mass of liquid water in terms of the size distribution function
$f(r,t)$ is $\bar{M}=\frac{4}{3}\pi\rho\int_{0}^\infty f(r, t) \,{\rm d}r$, which is
$\bar{M}=\int_{0}^\infty g(\ln r, t) \,{\rm d}\ln r$ in terms of $\rm{g}(\ln r, t)$.
Therefore, ${\rm g}(\ln{r},t)=(4\pi/3)\rho r^4 f(r,t)$.
Figure~\ref{m_lucky_gravRe100_cond0_coa_less} shows $g(\ln{r},t)$
calculated from the same simulations as
in Figure~\ref{f_lucky_gravRe100_cond0_coa}, where the collision is driven
by both gravity and turbulence ($\sigma_{\rm ini}=0$).
At $t=1\,$s, turbulence enhances the auto-conversion phase
as shown by the first peaks, when $10\,\mu$m-sized droplets collide.
The enhancement factor (amplitude of the peaks) scales almost linearly with $\bar{\epsilon}$.
This enhancement at the auto-conversion phase leads to faster growth of droplet with
increasing $\bar{\epsilon}$ at late times (i.e., $t=5$ s and $t=10$ s.).
Therefore, we see a faster growth of large droplets with increasing
$\bar{\epsilon}$ at late times virtually.
This is also consistent with the conclusion from
Figure~\ref{f_lucky_comp} that turbulence dominates
the collisional growth at the early stage of cloud droplets formation.
Additionally, this also implies that the turbulence enhancement effect is the most efficient
when the size of a colliding pairs is comparable, which is consistent
with previous findings
\citep{pinsky2007collisions, ayala_2008, chen2018turbulence}.

When gravity is included, the non-dimensional terminal velocity
$\rm{Sv}=v_g/u_{\eta}$, charactering the relative droplet
inertia and gravitational sedimentation, becomes important \citep{DBB12}
(Here we adopt $\rm Sv$ because it contains the information of particle size compared with
the Froude number of particles defined as
${\rm Fr}=|\mbox{\boldmath $\bm{g}$}|\tau_{\eta}/u_{\eta}$ \citep{Gustavsson14}.
$\rm{Sv}$ can be expressed as $\rm{Sv}=\rm{Fr\,St}$),
where $v_g=\tau_i |\bm{g}|$ is the terminal fall velocity and $u_{\eta}$
is the turbulent velocity at the Kolmogorov scale $\eta$.
It can also be interpreted as the
ratio of the Kolmogorov eddy turnover time and the time it takes for a
particle to sediment across the eddy. If the ratio is much larger than unity,
the particle will rapidly sediment through the eddy, thereby leading
to weak particle--eddy interaction.
On the other hand, if $\rm Sv$ is much smaller than unity,
sedimentation does not play a significant
role in reducing the time of particle--eddy interaction \citep{Ayala08}.
The distribution of $\rm Sv$ shows the same dependency on $\bar{\epsilon}$ as $f(r,t)$
in our simulations when $\sigma_{\rm ini}=0$.
as demonstrated in Figure~\ref{f_lucky_gravRe100_cond0_coa}.

\begin{figure}[t!]
\begin{center}
\includegraphics[width=0.5\textwidth]{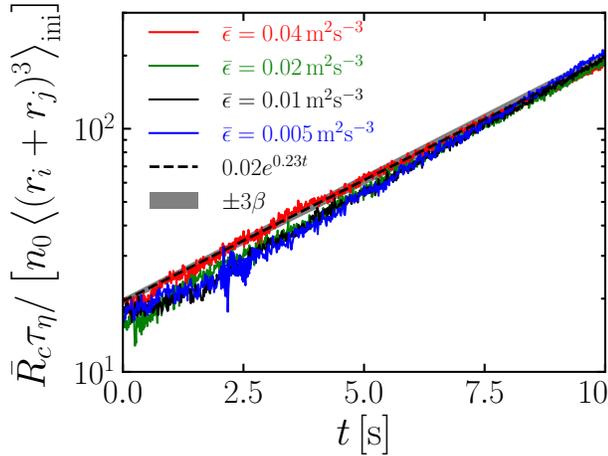}
\end{center}
\caption{Mean collision rate $\bar{R}_c$ for different $\bar{\epsilon}$.
Same simulations as in Figure~\ref{f_lucky_gravRe100_cond0_coa}.
Here $\beta$ is the standard deviation and $3\beta$ is the significance level.
}
\label{meanCaogulationRate}
\end{figure}

Further inspection of the mean collision rate $\bar{R}_c$
(Figure~\ref{meanCaogulationRate}) is consistent with
the above observations. More importantly, the normalized
$\bar{R}_c$ collapse onto each other and follow exponential growth.
The collapse reconciles our finding that the turbulence enhancement
at the auto-conversion phase scales with $\bar{\epsilon}$.
The exponential growth of $\bar{R}_c$
can be explained by the following theory of the continuous collision \citep{LV11}.
Given two droplets of very different sizes that collide with each other
due to gravity, the collision rate given by equation~\eqref{tauij1} is
\begin{equation}
	\bar{R_c}^{\rm continuous}=\pi\left(r_L+r_S\right)^2\left|\bm{V}_L
	-\bm{V}_{S}\right|.
\label{eq:continuousRc}
\end{equation}
Substituting equation~\eqref{eq:correction} into equation~\eqref{response_time},
and taking into account that
$\rm{Re}_i \sim V_ir_i$, we obtain $\tau_i \sim r_i^2/(V_i r_i)^{2/3}=r_i^{4/3}V_i^{-2/3}$.
When gravity dominates the motion of a droplet,
the droplet velocity is of the order of the terminal
fall velocity, $V_i=\tau_i g$, so that $V_i\sim r_i^{4/5}$.
The linear approximation for the velocity \citep{LV11} is now obtained by
replacing the exponent 4/5 with unity, such that equation~\eqref{eq:continuousRc}
simplifies to
\begin{equation}
	\bar{R_c}^{\rm continuous}\sim r_L^3
\label{eq:continuousRc2}
\end{equation}
when $r_L\gg r_S$.
The rate of mass increase, $d\rm{m_L}/dt$, is proportional to the collision rate.
Therefore, equation~\eqref{eq:continuousRc2} can also be expressed as
\begin{equation}
	\frac{d\rm{m_L}}{dt}\sim r_L^3.
\label{eq:continuousRc3}
\end{equation}
Combine equations~\eqref{eq:continuousRc2} and \eqref{eq:continuousRc3},
we can obtain the exponential growth of $\bar{R}_c$,
\begin{equation}
	\bar{R_c}^{\rm continuous}\sim \exp(\alpha t),
\label{eq:continuousRc4}
\end{equation}
where $\alpha$ is a constant.

The excellent agreement of $\bar{R}_c$ between our
simulation and the theory demonstrates that the continuous growth
theory is robust in representing the mean collision rate.
Even in the circumstance
that we detect the collision rate directly by counting each collision event
without any assumptions, such as that of large size differences, the linear approximation
for the velocity \citep{LV11}, and the absence of turbulence.
When $\bar{R}_c$ is normalized by $\tau_{\eta}$, the
curves representing different $\bar{\epsilon}$ collapse onto each other
(see Figure~\ref{meanCaogulationRate}).
This indicates that (1) gravity dominates the collisional growth; (2)
collision time scale is smaller than the Kolmogorov time scale; (3)
turbulence is responsible for generating few larger droplets so that
the gravitational collision can be triggered at the initial phase of raindrop
formation; (4) turbulence transport
provides a mean effect for collision as implicitly indicated in equation~\eqref{eq:ST}.

The mean collision rate $\bar{R_c}$ is an averaged description, which
ignores fluctuations.
The agreement between DNS results and
the theory of continuous collisions only suggests the mean effect
of turbulence on collisional growth. Collisional growth due to random
fluctuations in a dilute system (such as cloud system) was proposed already
by \citet{Telford1955}.
\citet{Kos05} further argued that Poisson fluctuations of the collision times of
settling droplets leads to a broad distribution of growth times that could
potentially explain the rapid onset of rain formation. This question is further
discussed by \citet{wilkinson16}.
Therefore, quantifying the role of fluctuations in the collision process by means of analyzing
the DNS data in the framework proposed by \citet{Kos05} and \citet{wilkinson16}
is desired.
Also, since our system is 100 times denser than the Earth's atmospheric
cloud system, it is interesting to investigate how the diluteness affects the role
of fluctuations on collisional growth.

\begin{figure*}[t!]
\begin{center}
\includegraphics[width=\textwidth]{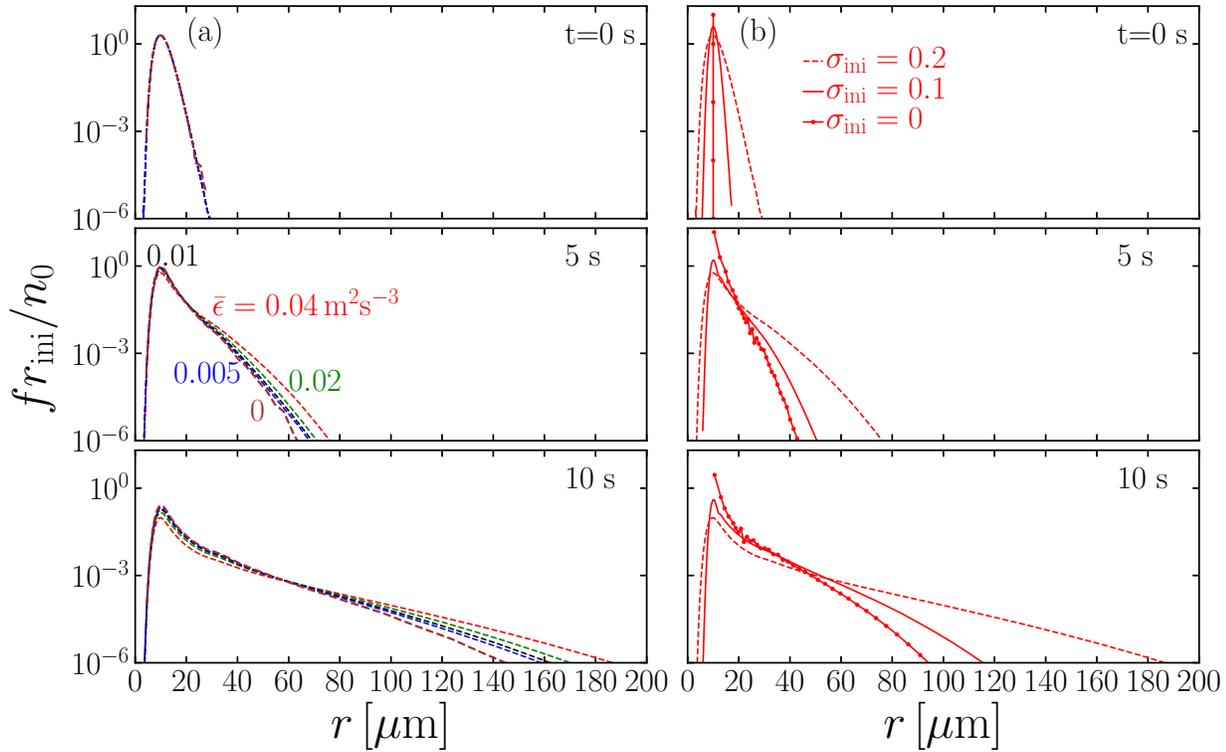}
\end{center}
\caption{Same simulations as in Figure~\ref{f_shima_grav0Re100_cond0_coa_comp} but with
gravity invoked. (a): time evolution of size distribution for different $\bar{\epsilon}$ with $\sigma_{\rm ini}=0.2$;
the brown dashed lines represent the size distribution due to gravity-generated collision.
(b): comparison among $\sigma_{\rm ini}=0$, 0.1, and 0.2 with $\bar{\epsilon}=0.04\,\rm{m}^2\rm{s}^{-3}$.
}
\label{f_shima_gravRe100_cond0_coa_comp}
\end{figure*}

In the atmospheric clouds, the size distribution of cloud droplets has a certain
width.
To investigate the collisional growth with a lognormal initial distribution
when there is both turbulence and gravity,
we use the same setup as in Figure~\ref{f_shima_grav0Re100_cond0_coa_comp}(b),
but with gravity included.
As shown in Figure~\ref{f_shima_gravRe100_cond0_coa_comp}(a),
the evolution of the droplet-size distribution
depends only weakly on the energy dissipation rate.
This again confirms the notion that gravity-generated collision is
more sensitive to the initial size difference
than the turbulence-generated collisions (see the $\sigma_{\rm ini}$ dependency
of the size distribution for turbulence-generated collision in appendix~\ref{app:width}).
To further illustrate
this, we plot the time evolution of the size distribution with different
initial widths; see Figure~\ref{f_shima_gravRe100_cond0_coa_comp}(b).
We also compare the present numerical simulations with
the idealized gravity-driven collision.
Figure~\ref{f_shima_gravRe100_cond0_coa_comp}(a) shows that
the tail of the size distribution becomes broadening as
$\bar{\epsilon}$ increases. For the case of $\bar{\epsilon}=0\,\rm{m}^2\rm{s}^{-3}$,
the tail reaches at about $142\,\mu$m at $t=10\,$s, while for the case of
$\bar{\epsilon}=0.04\,\rm{m}^2\rm{s}^{-3}$
the tail reaches at about $182\,\mu$m, resulting in an increase percentage
of $28\%$.
\citet{chen2018turbulence} found that the increasing percentage of the tail
is about $(45-30)/30=50\%$
at $t=6.5$ min (390 s). Our result reveals an increasing percentage of
$(75-62)/62\approx21\%$ at $t=5\,$s (equivalent
to $500\,$s considering that $n_0=10^{10}\rm{m}^{-3}$ is used in our simulations).
Since our initial size distribution and treatment of the collision efficiency
are different from the ones of \citet{chen2018turbulence},
we cannot compare our results with theirs quantitatively.
Nevertheless, our findings are consistent with the result of
\citet{chen2018turbulence} that
turbulence enhances the collisional growth of cloud droplets.
We recall that when $\sigma_{\rm ini}=0$, strong dependency of $f(r,t)$ on $\bar{\epsilon}$
is observed (see Figure~\ref{f_lucky_gravRe100_cond0_coa}).
This implies that the enhancement effect
of turbulence depends on the initial distribution of cloud droplets.

\section{Conclusions}

In the present study, we have addressed the problem
of turbulence effects on collisional growth of particles such as cloud droplets
in the case where coalescence is included.
We have investigated this effect
using a superparticle approximation for the droplet dynamics in
combination with high-resolution DNS of fully-developed turbulence.
The superparticle approach is about 10 times faster than
the direct Lagrangian-detected collision method at least.
In the absence of gravity, we find that the droplet-size distribution depends sensitively
on the mean energy-dissipation rate $\bar{\epsilon}$ at fixed ${\rm Re}_\lambda$,
which we have related to the $\bar{\epsilon}$-dependence of the mean collision rate.
We find that this rate increases
as $\bar{\epsilon}^{1/2}$ (except for the largest values of $\bar{\epsilon}$ simulated).
This is consistent with the Saffman-Turner collision model and its extensions.
A more detailed comparison with these
calculations is not possible at this point, because there is no prediction for the prefactors
in general.
The size distribution due to turbulence-generated collisions exhibits power law behavior
with a slope of $-3.7$
in the size range $10...40\,\mu$m, which is close
to the power law size distribution of interstellar dust grains. This indicates that
the power law size distribution may be universal \citep{mathis1977size}.
When gravity is invoked, the turbulence enhancement effect depends
on the width of the initial distribution $\sigma_{\rm ini}$.
The enhancement is the strongest when $\sigma_{\rm ini}=0$
and weak when $\sigma_{\rm ini}=0.2$.
For the case of $\sigma_{\rm ini}=0$, turbulence has an efficient
effect at the auto-conversion phase, which results in faster growth
at the late stage.
In atmospheric clouds, the
distribution of cloud droplets always has a certain width.
The role of turbulence for collisional growth should be handled
with caution.
To our knowledge, it is the first time that such detailed comparison between cases with
or without gravity is investigated when coalescence is invoked.

When collisions are driven by both turbulence and gravity,
we found that turbulence is crucial for driving the collision so that a few large cloud droplets
can be formed in the initial stage of raindrop formation.
Gravity takes over as the main driver for droplet collisions
when the radius of cloud droplets reaches the size of about $20\,\mu$m.
With combined turbulence and gravity, the time scale for reaching drizzle-sized droplets
is about 900 s, which is close to the time scale of the rapid warm
rain formation.
This time scale, however, is expected to be substantially changed
when turbulence-induced collision efficiency is taken into account.
The mean collision rate grows exponentially, which is consistent with
the theoretical prediction of the continuous growth even when turbulence is invoked.
The theory of continuous collisions is built upon the assumptions of huge size
differences, a linear drag force, and gravity-driven collisions.
The consistency between
our simulations and the theory suggests that the theory is robust
in representing the mean effect of turbulence.

The role of fluctuations for collisional growth
\citep{Telford1955,Kos05,wilkinson16} is not explicitly analyzed.
Therefore,
it is interesting to investigate how the diluteness affects the role
of fluctuations on collisional growth.
These will be presented in a separate paper.

Collisional growth of cloud droplets due to turbulence and gravity
is very sensitive to the tail of the initial size distribution.
As already discussed previously \citep{li17}, this problem is being alleviated
by considering the combined condensational and collisional growth.
Especially the condensational growth due to supersaturation fluctuations
may result in larger tails of the size distribution \citep{sardina2015, chandrakar2016}.
This is another subject of an ongoing separate study.

In the present paper, the collision efficiency
is assumed to be unity for simplicity.
In reality, the collision efficiency is not unity, but it can depend
on the droplet-droplet aero-hydrodynamics \citep{wang2005theoretical, wang2007effects,Wang_2009, chen2018turbulence}.
Using a unit collision efficiency overestimates the collision rate.
It would be useful to incorporate the collision efficiency in turbulence
invoking droplet-droplet aero-hydrodynamics,
but this has not yet been done.
This will need to be investigated in a separate study.

\acknowledgments

We thank Akshay Bhatnagar, Gregory Falkovich and Vladimir Zhdankin
for stimulating discussions.
We also thank the three anonymous reviewers for their constructive
suggestions and efforts to help improving our manuscript.
This work was supported through the FRINATEK grant 231444 under the
Research Council of Norway, SeRC, the Swedish Research Council grants
2012-5797 and 2013-03992,
the Israel Science Foundation governed by the
Israel Academy of Sciences (grant No. 1210/15),
the University of Colorado through its support of the
George Ellery Hale visiting faculty appointment,
and the grant ``Bottlenecks for particle growth in turbulent aerosols''
from the Knut and Alice Wallenberg Foundation, Dnr.\ KAW 2014.0048.
Gunilla Svensson also thanks the Wenner-Gren Foundation for their support."
The simulations were performed using resources provided by
the Swedish National Infrastructure for Computing (SNIC)
at the Royal Institute of Technology in Stockholm and
Chalmers Centre for Computational Science and Engineering (C3SE).
This work also benefited from computer resources made available through the
Norwegian NOTUR program, under award NN9405K.
The source code used for the simulations of this study, the {\sc Pencil Code},
is freely available on \url{https://github.com/pencil-code/}.
The input files as well as some of the output files of the simulations
listed in Table~\ref{Swarm_Rey} are available under
\url{http://www.nordita.org/~brandenb/projects/collision_turbulence/}.

\appendix

\section{Collision algorithm of the superparticle scheme}
\label{collision}

A detailed study of the superparticle approach is given in \citet{li17},
where its evaluation and advantages over the Eulerian approach are investigated.
Here, we briefly review the collision scheme used in the present study.
When two superparticles $i$ and $j$ residing in the same grid cell collide with
each other, the new masses of the particles in the two superparticles after collision
obey mass conservation and are given by
\begin{eqnarray}
\tilde{m}_i&=&m_{i}+m_j, \nonumber \\
\tilde{m}_j&=&m_j,
\end{eqnarray}
where $n_i$ and $n_j$ are the number densities of droplets in
superparticles $i$ and $j$, respectively. We assume $n_j>n_i$ without loss of generality.
Their new particle number densities are
\begin{eqnarray}
\tilde{n}_i&=&n_i, \nonumber \\
\tilde{n}_j&=&n_j-n_i.
\end{eqnarray}
The momenta of the particles in the two superparticles after collision are given by
\begin{eqnarray}
	\tilde{\bm{V}}_{i}\tilde{m}_{i}&=&\bm{V}_{i}m_{i}+\bm{V}_{j}m_j, \nonumber \\
	\tilde{\bm{V}}_{j}\tilde{m}_{j}&=&\bm{V}_{j}m_j.
\end{eqnarray}

\section{Effect of initial condition on collision in a turbulent environment}
\label{app:initialCondition}

\begin{figure}[t!]\begin{center}
\includegraphics[width=0.5\textwidth]{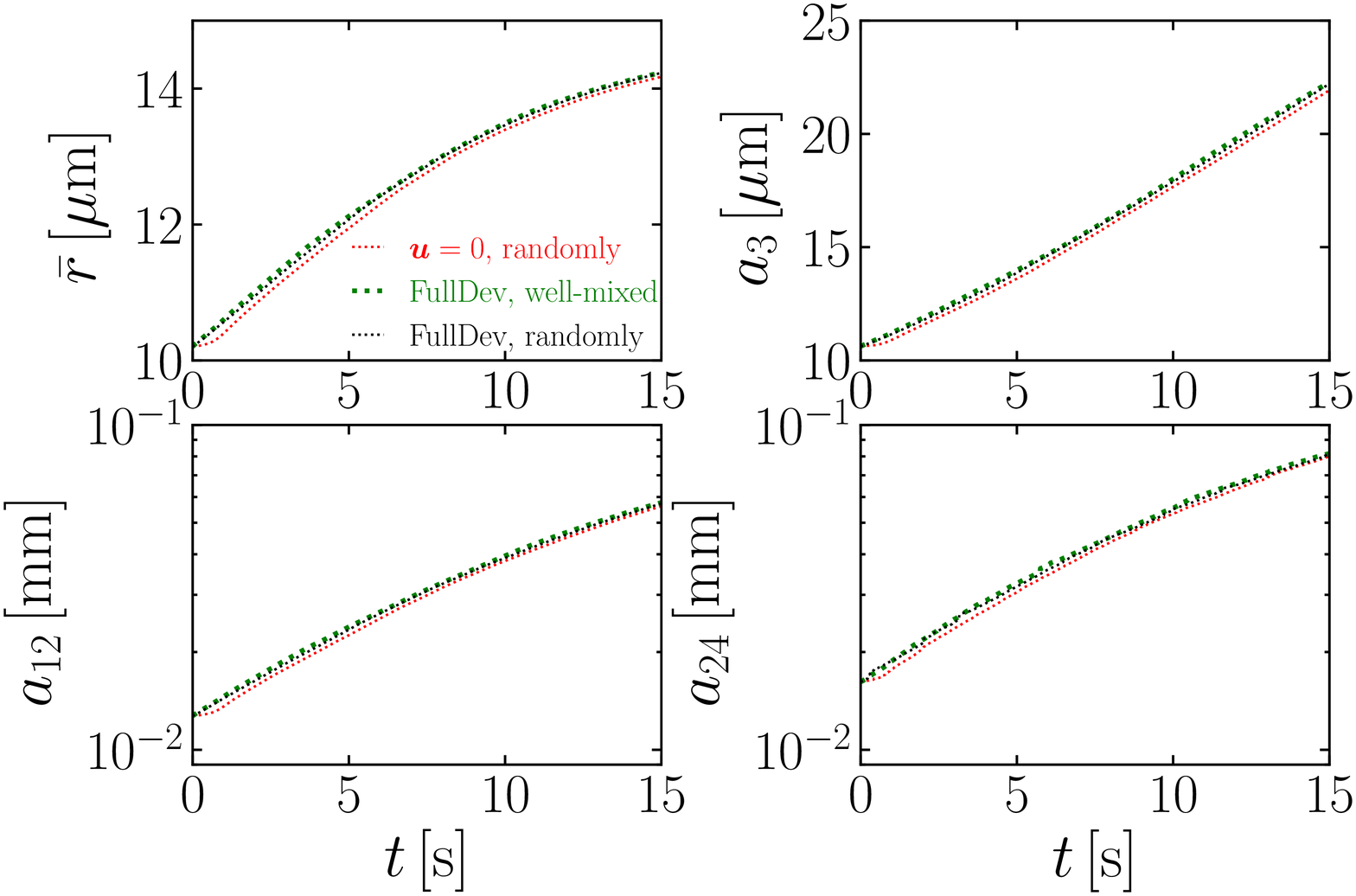}
\end{center}
\caption{Comparison of $a_\zeta$ for different initial conditions:
collision is triggered (1) (red curve)
in a randomly distributed superparticle field and the velocity of
the flow is zero, (2) (black curve) in a well-mixed particle field and the velocity
of the flow is zero, and (3) (green curve) in a well-mixed particle field and the
turbulence is well-developed.
Gravity is omitted here. $L=0.25\,\rm{m}$. The initial size distribution is given
by equation~\eqref{init_dist} with $r_{\rm ini}=10\,\mu$m and $\sigma_{\rm{ini}}=0.2$.
The number of mesh grid points is $128^3$. $f_0=0.02$. $N_p/10^6=8.4$.
These result in $\mbox{\rm Re}_{\lambda}$ = 100 and
$\bar{\epsilon}$ = $0.03\rm{m}^2\rm{s}^{-3}$.
}
\label{moment_con0_coa_grav0_tur_comp_reinitialize}
\end{figure}

The initial conditions are important for the collision. We tested
three different initial conditions.
Collision is triggered (1)
in a randomly distributed superparticle field and the velocity of
the flow is zero, (2) in a well-mixed particle field and the velocity
of the flow is zero, and (3) in a well-mixed particle field and the
turbulence is well-developed.
To compare the time evolution of the size distribution for these three
cases, we first define the normalized moments of the size distribution \citep{li17},
\begin{equation}
	a_\zeta=\left(\left.\int_{0}^\infty f\,r^\zeta \,{\rm d}r\right/\,\int_{0}^\infty f \,{\rm d}r \right)^{1/\zeta}
\label{azeta}
\end{equation}
where $\zeta$ is a positive integer.
The mean radius $\overline{r}$ is given by $a_1$,
the maximum radius is $\max(r)=a_\infty$, and
the droplet mass is proportional to the third power of $a_3$.
$a_{\zeta}$ can characterize the size distribution with simpler diagnostics.
Figure~\ref{moment_con0_coa_grav0_tur_comp_reinitialize} shows $a_\zeta$
for the three different initial conditions. It is obvious
that the time evolution of $a_\zeta$ is independent from initial conditions.
This can sufficiently save computational time.

\section{Statistical convergence of the number of superparticles per grid cell in a turbulent environment}
\label{app:Np/Ngrid}

\begin{figure}[t!]\begin{center}
\includegraphics[width=0.5\textwidth]{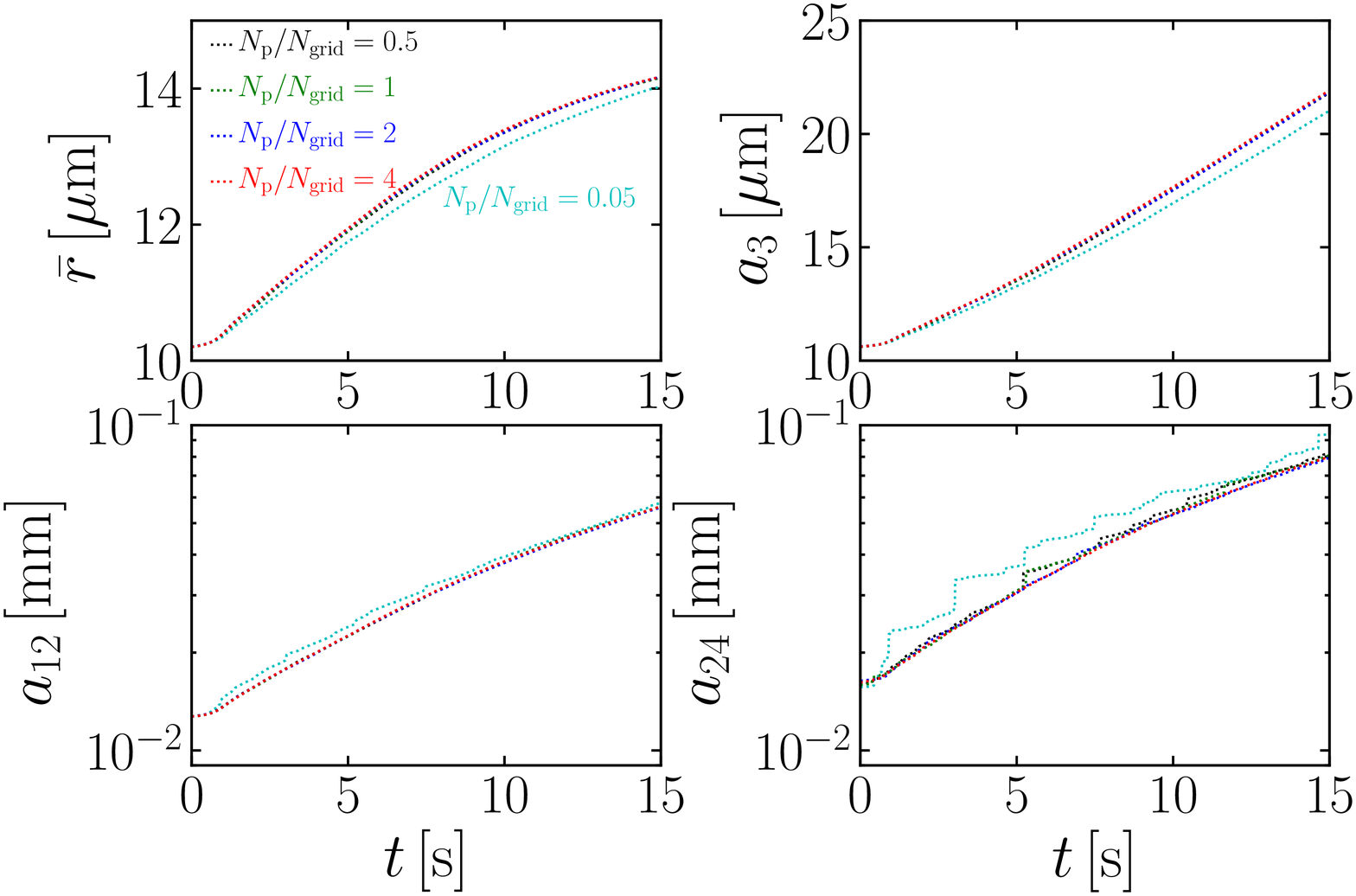}
\end{center}
\caption{Comparison of $a_\zeta$ for different
$N_{\rm p}/N_{\rm grid}$.
Here $N_{\rm grid}=128^3$ is fixed.
$N_{\rm p}/N_{\rm grid}$ even converges at $0.05$.
Same simulations as in Figure~\ref{moment_con0_coa_grav0_tur_comp_reinitialize}
but with different $N_{\rm p}$.
}
\label{moment_tur_npar}
\end{figure}

As discussed in Section~\ref{NumericalSetup}, simulations with massive number of
superparticles is computationally costing.
In \citet{li17}, we found that the initial $N_{\rm p}/N_{\rm grid}$ converges at 4 when the
collision is driven by gravity without turbulence.
In the present study, $N_{\rm grid}=512^3$. $N_{\rm p}/N_{\rm grid}=4$ will
result in $N_{\rm p}=4\times512^3=536870912$, which will be very computationally
demanding.
This motivates us to re-study the convergence of $N_{\rm p}/N_{\rm grid}$
in high Reynolds number turbulence case instead of carrying the convergence
study from pure gravity case to the turbulence case. As shown in
Figure~\ref{moment_tur_npar}, $N_{\rm p}/N_{\rm grid}$ converges at $0.5$.
This could be due to the fact that turbulence transports particles sufficiently.

\section{Convergence of the initial width $\sigma_{\rm ini}$ in a turbulent environment with gravity}
\label{app:width}
\begin{figure}[t!]
\begin{center}
\includegraphics[width=0.5\textwidth]{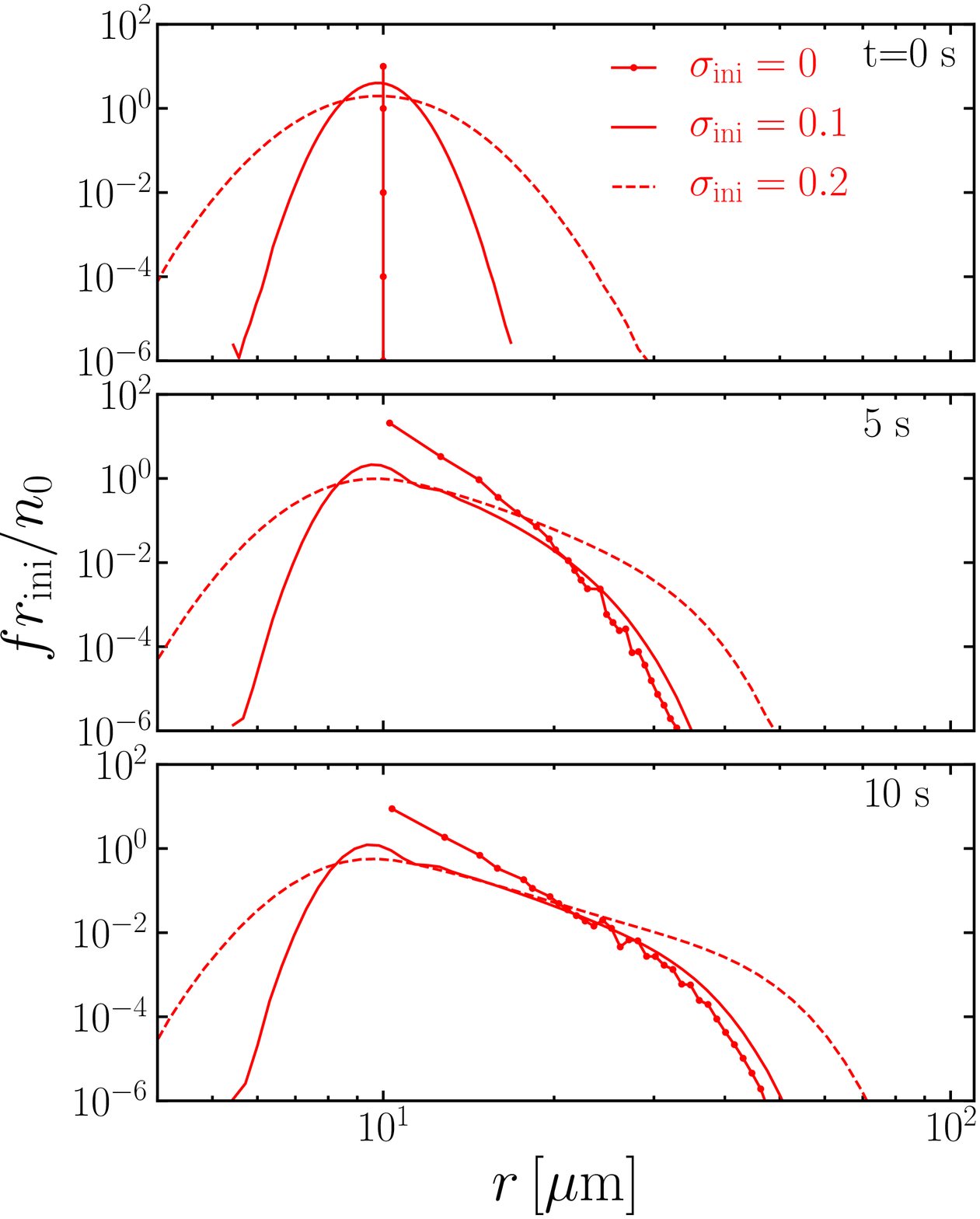}
\end{center}
\caption{Convergence of $\sigma_{\rm ini}$.
Collision is driven solely by turbulence;
see Run B for simulation details.}
\label{f_shima_grav0Re100_cond0_coa_sig_comp}
\end{figure}

\begin{figure}[t!]
\begin{center}
\includegraphics[width=0.5\textwidth]{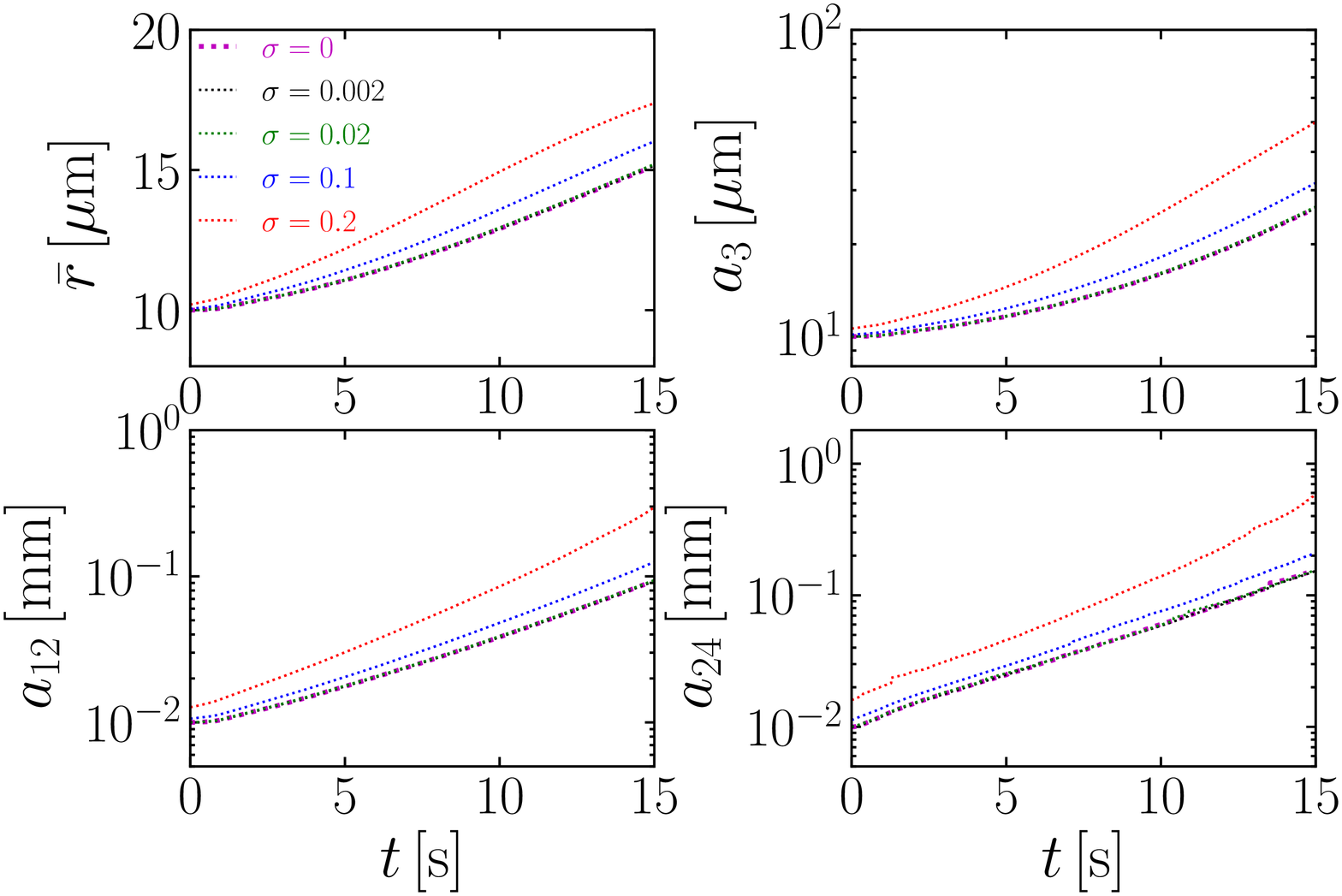}
\end{center}
\caption{Convergence of $\sigma_{\rm ini}$. Collision is driven by combined turbulence and
gravity; see Run B for simulation details.}
\label{moments_sig_comp_turb_grav}
\end{figure}

We first check the convergence of $\sigma_{\rm ini}$
when collisions are driven solely by turbulence.
As shown in Figure~\ref{f_shima_grav0Re100_cond0_coa_sig_comp}, the time evolution
of the size distribution almost converges at $\sigma_{\rm ini}=0.1$.
Compared with the case where the collision is driven by combined turbulence
and gravity as demonstrated in Figure~\ref{f_shima_gravRe100_cond0_coa_comp},
the tail of the size distribution is less sensitive to $\sigma_{\rm ini}$.
Next,
we investigate the convergence of the width $\sigma_{\rm ini}$ in equation~\eqref{init_dist}
in a combined turbulence and gravity environment.
Figure~\ref{moments_sig_comp_turb_grav} shows that $\sigma_{\rm ini}$ converges at
$0.02$.
However, as we have discussed in Section~\ref{gt}, we choose $\sigma_{\rm ini}=0$ for
the combined turbulence and gravity case.

\section{Distribution of the Stokes number}
\label{app:StokesNumner}

\begin{figure}[t!]
\begin{center}
\includegraphics[width=0.49\textwidth]{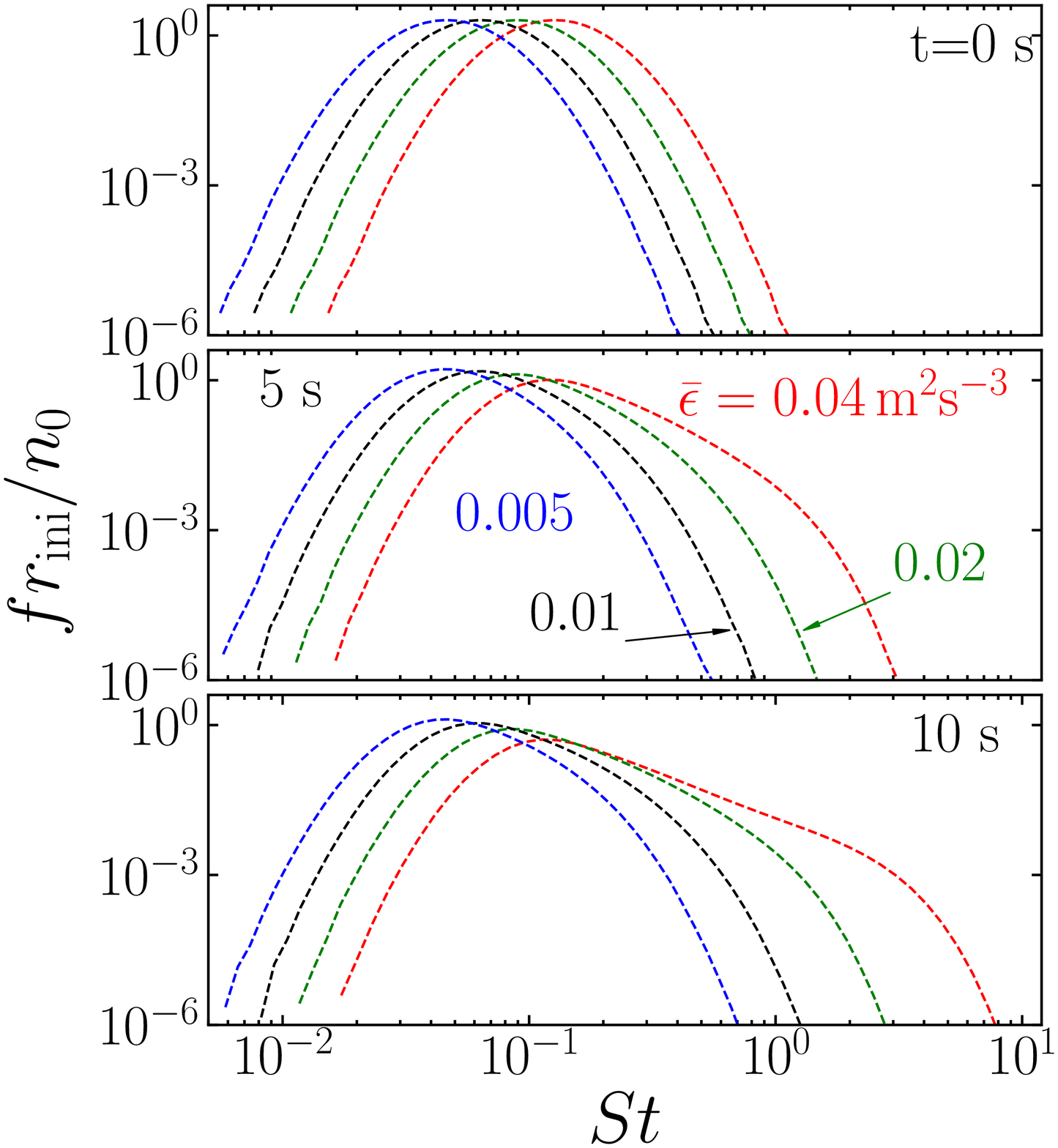}
\end{center}
\caption{Time evolution of the Stokes number distribution $St(r,t)$
for different $\bar{\epsilon}$ with fixed $Re_{\lambda}$.
Same simulations as in Figure~\ref{f_shima_grav0Re100_cond0_coa_comp}.}
\label{Stokes_grav0Re100_cond0_coa_comp}
\end{figure}

Figure~\ref{Stokes_grav0Re100_cond0_coa_comp} shows the distribution function of Stokes numbers for the same
simulations as in Figure~\ref{f_shima_grav0Re100_cond0_coa_comp}. Initially, the distribution of Stokes number
shifts to the right with increasing $\bar{\epsilon}$,
which will trigger stronger collisional growth. At later times, when $\bar{\epsilon}$
increases from $0.005$ to $0.04\,\rm{m}^2 \rm{s}^{−3}$, the tail of the Stokes number distribution
increases by more than an order of magnitude, which leads to an extension of about three orders
of magnitude at $t=10$ s. This indicates that the collisional growth rate strongly depends on the Stokes
number. Increasing $\bar{\epsilon}$ results in a larger range of variations in the value of the Stokes number,
thus enhancing the collisional growth.

\section{Rescaling of time based on the initial number density}
\label{rescaling}

\begin{figure}[t!]\begin{center}
\includegraphics[width=0.49\textwidth]{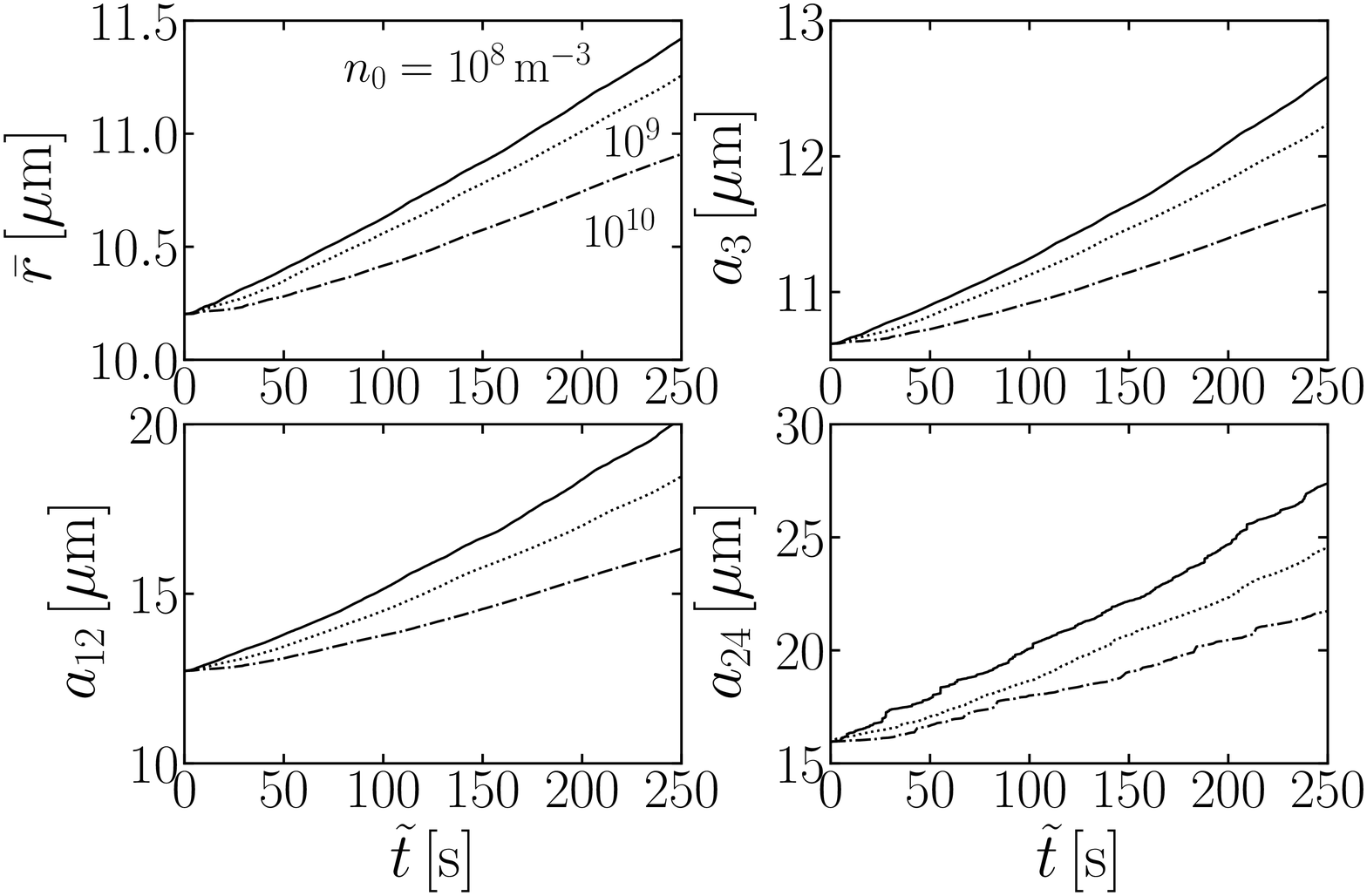}
\end{center}
\caption{Accuracy of the rescaling procedure of time in 2-D turbulence.
The size of the computational domain is $L=0.25$ m,
the number of grid cells is $N_{\rm grid}=512^2$,
the viscosity is $\nu=5\times10^{-4} \, \rm{m}^2{\rm s}^{-1}$, which gives a
Taylor micro-scale Reynolds number of $Re_{\lambda}\approx106$, and an energy
dissipation rate of $\epsilon = 0.1 \,\rm{m}^2{\rm{s}}^{-3}$.
The initial droplet size distribution is log-normal with $r_{\rm ini}=10\mu$m
and $\sigma=0.2$. The number of the superparticles is $N_{\rm s}=1.2\times10^6$
and the number of the grid cell is $N_{\rm grid}=512^2$. Each simulation was conducted
on 512 CPUs.}
\label{moment_con0_coa_grav_rescale_2D_SP}
\end{figure}

As explained at the end of section~\ref{NumericalSetup},
simulating the collision-coalescence process of raindrop formation
over realistic time scales is computationally demanding, so we adopt an initial
number density of $n_0=10^{10}\,\rm{m}^{-3}$ and rescale the simulation time to
$\tilde{t}=t(n_0/n_{\rm ref})$. We check $a_\zeta$ for the rescaling in 2-D
turbulence.
As shown in Figure~\ref{moment_con0_coa_grav_rescale_2D_SP}, larger $n_0$
result in smaller $a_\zeta$. However, the difference in $a_{24}$
is only about $5\mu$m at $\tilde{t}=250$ s. This means that using
$n_0=10^{10}\,\rm{m}^{-3}$ does reasonably well represent the collisional growth.

\bibliographystyle{ametsoc2014.bst}
\bibliography{database}

\end{document}